\documentclass[twocolumn,floatfix,eqsecnum,rmp]{revtex4}
\usepackage{graphicx}
\usepackage{amsmath}
\usepackage{bm}
\bibpunct{}{}{,}{s}{}{\textsuperscript{,}}
\newcommand{\be}{\begin{equation}}
\newcommand{\ee}{\end{equation}}

\begin{document}
\title{Coulomb-Blockade Oscillations in Semiconductor Nanostructures}
\author{H. van Houten, C. W. J. Beenakker, and A. A. M. Staring}
\affiliation{Philips Research Laboratories, 5600 JA Eindhoven, The Netherlands}
\begin{abstract}
{\tt Published in {\em Single Charge Tunneling}, edited by H. Grabert and M. H. Devoret,\\
NATO ASI Series B294 (Plenum, New York, 1992).}
\end{abstract}
\maketitle

\tableofcontents

\section{\label{sec1} Introduction}

\subsection{\label{sec1.1} Preface}

\begin{figure}
\centerline{\includegraphics[width=8cm]{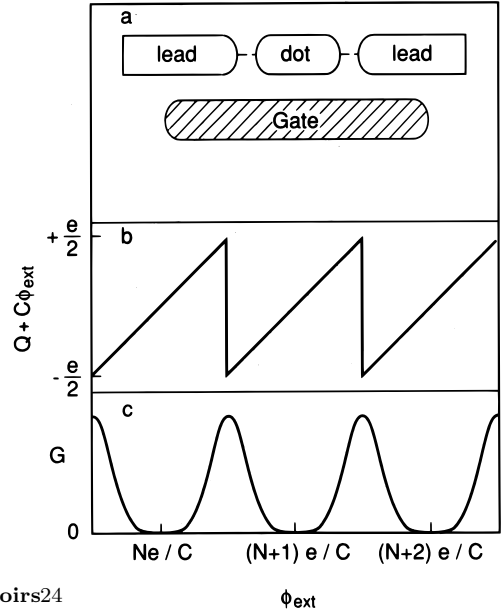}}
\caption{
(a) Schematic illustration of a confined region (dot) which is weakly coupled by tunnel
barriers to two leads. (b) Because the charge $Q=-Ne$ on the dot can only change by multiples of
the elementary charge $e$, a charge imbalance $Q+C\phi_{\rm{ext}}$ arises between the dot and the leads. This
charge imbalance oscillates in a saw-tooth pattern as the electrostatic potential $\phi_{\rm{ext}}$ is varied ($\phi_{\rm{ext}}$ is
proportional to the gate voltage). (c) Tunneling is possible only near the charge-degeneracy points
of the saw-tooth, so that the conductance $G$ exhibits oscillations. These are the ``Coulomb-blockade
oscillations''.
\label{fig1}
}
\end{figure}

Coulomb-blockade oscillations of the conductance are a manifestation of single-electron tunneling through a system of two tunnel junctions in series (see Fig.\ \ref{fig1}).\cite{ref1,ref2,ref3,ref4,ref5}
The conductance oscillations occur as the voltage on a nearby gate electrode is varied.
The number $N$ of conduction electrons on an island (or dot) between two tunnel barriers is an integer, so that
the charge $Q=-Ne$ on the island can only change by discrete amounts $e$. In contrast,
the electrostatic potential difference of island and leads changes continuously as the electrostatic potential $\phi_{\rm{ext}}$ due to the gate is varied. This gives rise to a net charge imbalance
$C\phi_{\rm{ext}}-Ne$ between the island and the leads, which oscillates in a saw-tooth pattern with
gate voltage ($C$ is the mutual capacitance of island and leads). Tunneling is blocked at
low temperatures, except near the degeneracy points of the saw-tooth, where the charge
imbalance jumps from $+e/2$ to $-e/2$. At these points the Coulomb blockade of tunneling is lifted and the conductance exhibits a peak. In metals these ``Coulomb-blockade oscillations'' are essentially a classical phenomenon.\cite{ref6,ref7}
Because the energy level separation $\Delta E$ in the island is much smaller than the
thermal energy $k_{\mathrm{B}}T$, the energy spectrum may be treated as a continuum. Furthermore,
provided that the tunnel resistance is large compared to the resistance quantum $h/e^{2}$,
the number $N$ of electrons on the island may be treated as a sharply defined classical
variable.

Coulomb-blockade oscillations can now also be studied in semiconductor nanostructures, which have a discrete energy spectrum. Semiconductor nanostructures are fabricated by lateral confinement of the two-dimensional electron gas $(2\rm{DEG})$ in Si-inversion layers, or in GaAs-AlGaAs heterostructures. At low temperatures, the conduction electrons
in these systems move over large distances (many $\mu \mathrm{m}$) without being scattered
inelastically, so that phase coherence is maintained. Residual elastic scattering by impurities or off the electrostatically defined sample boundaries does not destroy this phase
coherence. The Fermi wavelength $\lambda_{\mathrm{F}}\sim 50\,\mathrm{n}\mathrm{m}$ in these systems is comparable to the
size of the smallest structures that can now be made using electron-beam lithography.
This has led to the discovery of a variety of quantum size effects in the ballistic transport regime. These effects may be adequately understood without considering electron-electron interactions.\cite{ref8}

\begin{figure}
\centerline{\includegraphics[width=6cm]{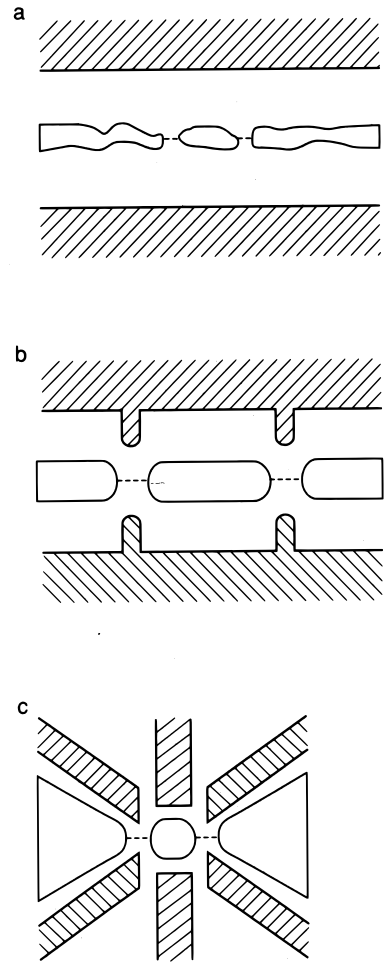}}
\caption{
Schematic top-view of three semiconductor nanostructures exhibiting Coulomb-blockade
oscillations. Hatched regions denote gates, electron gas regions are shaded. Dashed lines indicate
tunneling paths. (a) Disordered quantum wire with a single conductance limiting segment. (b) Quantum
dot in a narrow channel. (c) Quantum dot between wide regions with separate sets of gates to modulate
the tunnel barriers, and to vary the external potential of the dot.
\label{fig2}
}
\end{figure}

The first type of semiconductor nanostructure found to exhibit Coulomb-blockade oscillations is a narrow disordered wire, defined by a split-gate technique.\cite{ref9,ref10,ref11,ref12,ref13,ref14} As
shown in Fig.\ \ref{fig2}a, such a quantum wire may break up into disconnected segments if it
is close to pinch-off. Conduction at low temperatures proceeds by tunneling through
the barriers delimiting a segment, which plays the role of the central island in Fig.\ \ref{fig1}.
The dominant oscillations in a wire typically have a well-defined periodicity, indicating
that a single segment limits the conductance. Nevertheless, the presence of additional
segments may give rise to multiple periodicities and to beating effects.

The second type of nanostructure exhibiting Coulomb-blockade oscillations is a small
artificially confined region in a 2DEG (a quantum dot), connected by tunnel barriers either to narrow leads (Fig.\ \ref{fig2}b),\cite{ref15,ref16} or to wide electron reservoirs (Fig.\ \ref{fig2}c).\cite{ref17}
The distinction between these two types of nanostructures is not fundamental, since a
segment of a quantum wire delimited by two particularly strong scattering centers can
be seen as a naturally formed quantum dot. Both types of structure are of interest:

Whereas artificially defined quantum dots are more suited to a study of the effect under relatively well-controlled conditions, the significance of the phenomenon of {\it periodic\/}
conductance oscillations in {\it disordered\/} quantum wires lies in its bearing on the general
problem of transport in disordered systems. It contradicts the presumed ubiquity of
random conductance fluctuations in mesoscopic systems, and directly demonstrates the
predominant role of electrostatic interactions in a disordered conductor.\cite{ref18}

In a typical experiment, the segment of the wire, or the quantum dot, contains $ N\sim 100$ electrons, with an average energy level separation $\Delta E\sim 0.2\,\mathrm{m}\mathrm{e}\mathrm{V}$. At temperatures
below a few Kelvin, the level spacing $\Delta E$ exceeds the thermal energy $k_{\mathrm{B}}T$, so that
transport through the quantum dot proceeds by resonant tunneling. Resonant tunneling
can by itself also lead to conductance oscillations as a function of gate voltage or Fermi
energy. The interplay of resonant tunneling and the Coulomb blockade occurs when
$\Delta E$ and the charging energy $e^{2}/C$ are of comparable magnitude (which is the case
experimentally, where $e^{2}/C\sim 1\,\mathrm{m}\mathrm{e}\mathrm{V}$). This chapter reviews our current understanding
of this interplay in semiconductor nanostructures. After a brief introduction to the
properties of a 2DEG (based on Ref.\ \cite{ref8}) we present in Sec.\ \ref{sec2} a discussion of the key results
of a linear response theory for Coulomb-blockade oscillations in a quantum dot.\cite{ref19,ref20} In
Sec.\ \ref{sec3} we review experimental results on quantum dots\cite{ref15,ref16,ref17} and disordered quantum
wires\cite{ref9,ref10,ref11,ref12,ref13,ref14} in the absence of a magnetic field, and discuss to what extent they are now understood.

Kastner and collaborators\cite{ref9,ref10,ref15,ref21} originally suggested that the conductance
oscillations which they observed were due to the formation of a charge density wave or
``Wigner crystal''. They inferred from a model due to Larkin and Lee,\cite{ref22}, and Lee and
Rice,\cite{ref23} that the conductance would be thermally activated because of the pinning
of the charge density wave by impurities in the narrow channel. The activation energy
would be determined by the most strongly pinned segment in the channel, and periodic
oscillations in the conductance as a function of gate voltage or electron density would
reflect the condition that an integer number of electrons is contained between the two
impurities delimiting that specific segment. A Wigner crystal is a manifestation of long-range order neglected in the theory of Coulomb-blockade oscillations. In a quantum wire
with weak disorder (no tunnel barriers), a Wigner crystal may well be an appropriate
description of the ground state.\cite{ref24} The point of view adopted in this chapter, following
Ref.\ \cite{ref25}, is that the Coulomb blockade model is adequate for the present experiments
in systems with artificial or natural tunnel barriers. We limit ourselves to a discussion
of that model, and refer the reader to Ref.\ \cite{ref11} for an exposition of the alternative point
of view of Kastner and collaborators.

The Coulomb blockade and Wigner crystal models have in common that electron-electron interactions play a central role. In contrast, some authors have argued that
resonant tunneling of non-interacting electrons can by itself explain the observed conductance oscillations.\cite{ref26,ref27} We stress that one cannot discriminate between these
two models on the basis of the periodicity of the oscillations. Conductance oscillations
due to resonant tunneling through non-degenerate levels as well as Coulomb-blockade
oscillations both have a periodicity corresponding to the addition of a single electron
to the confined region. Other considerations (notably the absence of spin-splitting of
the peaks in a magnetic field, and the large activation energy --- by far exceeding $\Delta E$)
are necessary to demonstrate the inadequacy of a model based on resonant tunneling of
non-interacting electrons.

Semiconductor nanostructures offer the additional intriguing possibility to study
single-electron tunneling in the quantum Hall effect regime. This is the subject of Sec.\ \ref{sec4}.
In this regime of a strong magnetic field, the one-electron states are extended along
equipotential contours.\cite{ref8} The contours of subsequent states within the same Landau
level enclose one extra flux quantum $h/e$. States at the Fermi level are edge states circulating along the circumference of the quantum dot. If charging effects are negligible
oscillations in the conductance of the dot are observed as a function of gate voltage or
magnetic field, due to resonant tunneling through circulating edge states.\cite{ref28} This is a
manifestation of the Aharonov-Bohm effect, normally associated with magnetoconductance oscillations in a ring, rather than a dot. Circulating edge states, however, make
the dot behave effectively as a ring\cite{ref29} --- at least for non-interacting electrons. As we
will discuss, the single-electron charging energy can cause a``Coulomb blockade'' of the
Aharonov-Bohm effect in a quantum dot.\cite{ref30,ref31} The magnetoconductance oscillations
are suppressed when $e^{2}/C$ becomes comparable to the Landau level separation $\hbar\omega_{\mathrm{c}}$ (with
$\omega_{\mathrm{c}}=eB/m$). However, the periodic oscillations as a function of gate voltage remain.
This difference illustrates how in the presence of charging effects magnetic and electrostatic fields play fundamentally different roles,\cite{ref12} in contrast to the equivalent roles
played in the diffusive or ballistic transport regimes.\footnote{Examples of this equivalence are the fluctuations in the conductance as a function of gate voltage
or magnetic field due to quantum interference, and the sequence of quantized conductance plateaux
(at integer multiples of $e^{2}/h$) as a result of magnetic or electrostatic depopulation of one-dimensional
subbands.\cite{ref8}}
An additional topic covered in
Sec.\ \ref{sec4} is the effect of a magnetic field on the amplitude and position of the oscillations,
from which detailed information can be obtained on the one-electron energy spectrum
of the quantum dot.\cite{ref32}

In this chapter we consider the Coulomb-blockade oscillations in zero magnetic field
and in the {\it integer\/} quantum Hall effect regime. The generalization to the {\it fractional\/}
quantum Hall effect is still an open problem, at least experimentally. Some theoretical
considerations have been given,\cite{ref33} but will not be considered here. We limit ourselves
to the linear response regime, and do not discuss the non-linear current-voltage characteristics.\cite{ref34,ref35} 
In metallic tunnel junctions with very different tunnel rates through
the two barriers one finds steps in the current as a function of source-drain voltage.\cite{ref1,ref2}
This ``Coulomb staircase'' has recently also been observed
in a quantum dot.\cite{ref36} A third limitation is to stationary transport phenomena, so that
we do not consider the effects of radio-frequency modulation of the source-drain or gate
voltages. A new development in metals is the realization of a``turnstile clocking'' of the
current through an array of junctions at a value $ef$, with $f$ the frequency of the modulation of the voltage on a gate.\cite{ref37,ref38} These effects very recently also been observed in a quantum dot.\cite{ref36} Concerning the types of sample,
we limit ourselves to quantum dots and wires defined by a split-gate in a two-dimensional
electron gas. Quantum dots may also be defined by etching a pillar out of a quantum
well.\cite{ref39,ref40} Such ``vertical'' structures have the advantage over the planar structures
considered here that the thickness and height of the potential barriers separating the
quantum dot from the leads can be tailored to a great precision during the epitaxial
growth. A disadvantage is that it is more difficult to change the carrier density in the
dot by means of a gate electrode.\cite{ref41} In the planar structures based on a 2DEG not
only the electron density, but also the geometry can be varied continuously using gates.

\subsection{\label{sec1.2} Basic properties of semiconductor nanostructures}

Electrons in a two-dimensional electron gas (2DEG) are constrained to move in a
plane, due to a strong electrostatic confinement at the interface between two semiconductor layers (in the case of a GaAs-AlGaAs heterostructure), or at the interface between a
semiconductor and an insulator (in the case of a Si-inversion layer, where the insulator is
$\mathrm{S}\mathrm{i}\mathrm{O}_{2})$. The areal density $n_{\mathrm{s}}$ may be varied continuously by changing the voltage on a gate
electrode deposited on the top semiconductor layer (in which case isolation is provided
automatically by a Schottky barrier) or on the insulator. The gate voltage is defined
with respect to an ohmic contact to the 2DEG. The density under a gate electrode of
large area changes linearly with the electrostatic potential of the gate $\phi_{\mathrm{gate}}$, according to
the plate capacitor formula
\be
\delta n_{\mathrm{s}}=\frac{\epsilon}{ed}\delta\phi_{\mathrm{gate}}, \label{eq1}
\ee
where $\epsilon$ is the dielectric constant of the material of thickness $d$ between gate and 2DEG.
For GaAs $\epsilon=13\,\epsilon_{0}$, whereas $\mathrm{S}\mathrm{i}\mathrm{O}_{2}$ has $\epsilon=3.9\,\epsilon_{0}$.

A unique feature of a 2DEG is that it can be given any desired shape using lithographic techniques. The shape is defined by etching a pattern (resulting in a permanent
removal of the electron gas), or by electrostatic depletion using a patterned gate electrode
(which is reversible). A local (partial) depletion of the 2DEG under a gate is associated
with a local increase of the electrostatic potential, relative to the undepleted region. At
the boundaries of the gate a potential step is thus induced in the 2DEG. The potential
step is smooth, because of the large depletion length (of the order of 100 nm for a step
height of 10 meV). This large depletion length is at the basis of the split-gate technique,
used to define narrow channels of variable width with smooth boundaries.

The energy of non-interacting conduction electrons in an unbounded 2DEG is given
by
\be
E(k)=\frac{\hbar^{2}k^{2}}{2m}, \label{eq2}
\ee
as a function of momentum $\hbar k$. The effective mass $m$ is considerably smaller than the
free electron mass $m_{\mathrm{e}}$ as a result of interactions with the lattice potential (for GaAs
$m=0.067\,m_{\mathrm{e}}$, for Si $m=0.19\,m_{\mathrm{e}}$, both for the $(100)$ crystal plane). The density of
states $\rho_{2\mathrm{D}}(E)\equiv \mathrm{d}n(E)/\mathrm{d}E$ is the derivative of the number of electronic states $n(E)$ (per
unit surface area) with energy smaller than $E$. In $k$-space, these states fill a circle of
area $A=2\pi mE/\hbar^{2}$ [according to Eq.\ (\ref{eq2})], containing a number $g_{\mathrm{s}}g_{\mathrm{v}}A/(2\pi)^{2}$ of states.
The factors $g_{\mathrm{s}}$ and $g_{\mathrm{v}}$ account for the spin and valley-degeneracy, respectively (in GaAs
$g_{\mathrm{v}}=1$, in Si $g_{\mathrm{v}}=2$; $g_{\mathrm{s}}=2$ in zero magnetic field). One thus finds $n(E)=g_{\mathrm{s}}g_{\mathrm{v}}mE/2\pi\hbar^{2}$,
so that the density of states per unit area,
\be
\rho_{2\mathrm{D}}=g_{\mathrm{s}}g_{\mathrm{v}}\frac{m}{2\pi\hbar^{2}},\label{eq3}
\ee
is {\it independent\/} of the energy. In equilibrium, the states are occupied according to the
Fermi-Dirac distribution function
\be
f(E-E_{\mathrm{F}})=\left[1+\exp(\frac{E-E_{\mathrm{F}}}{k_{\mathrm{B}}T})\right]^{-1}.\label{eq4}
\ee
At low temperatures $k_{\mathrm{B}}T\ll E_{\mathrm{F}}$, the Fermi energy (or chemical potential) $E_{\mathrm{F}}$ of a 2DEG
is thus directly proportional to its sheet density $n_{\mathrm{s}}$, according to
\be
E_{\mathrm{F}}=n_{\mathrm{s}}/\rho_{2\mathrm{D}}.\label{eq5}
\ee
The Fermi wave number $k_{\mathrm{F}}\equiv(2mE_{\mathrm{F}}/\hbar^{2})^{1/2}$ is related to the density by $k_{\mathrm{F}}=$
$(4\pi n_{\mathrm{s}}/g_{\mathrm{s}}g_{\mathrm{v}})^{1/2}$. Typically, $E_{\mathrm{F}}\sim 10\,\mathrm{m}\mathrm{e}\mathrm{V}$, so that the Fermi wavelength $\lambda_{\mathrm{F}}\equiv 2\pi/k_{\mathrm{F}}\sim 50\,\mathrm{n}\mathrm{m}$.

If the 2DEG is confined laterally to a narrow channel, then Eq.\ (2) only represents
the kinetic energy from the free motion (with momentum $hk$) {\it parallel} to the channel
axis. Because of the lateral confinement, the conduction band is split itself into a series
of one-dimensional (1D) subbands, with band bottoms at $E_{n}$, $n=1,2, \ldots$. The total
energy $E_{n}(k)$ of an electron in the {\it n}-th 1D subband is given by
\be
E_{n}(k)=E_{n}+\frac{\hbar^{2}k^{2}}{2m},\label{eq6}
\ee
in zero magnetic field. Two frequently used potentials to model analytically the lateral confinement are the square well potential (of width $W$), and the parabolic potential well (described by $V(x)=\frac{1}{2}m\omega_{0}^{2}x^{2}$). The confinement levels are given by
$E_{n}=(n\pi\hbar)^{2}/2mW^{2}$, and $E_{n}=(n-\frac{1}{2})\hbar\omega_{0}$, respectively.

Transport through a very short quantum wire (of length $L\sim 100\,{\rm nm}$, much shorter
than the mean free path) is perfectly ballistic. When such a short and narrow wire forms
a constriction between two wide electron gas reservoirs, one speaks of a quantum point
contact.\cite{ref42} The conductance $G$ of a quantum point contact is quantized in units of
$2e^{2}/h$.\cite{ref42,ref44} This effect requires a unit transmission probability for all of the occupied
1D subbands in the point contact, each of which then contributes $2e^{2}/h$ to the conductance (for $g_{\mathrm{s}}g_{\mathrm{v}}=2$). Potential fluctuations due to the random distribution of ionized
donors have so far precluded any observation of the conductance quantization in longer
quantum wires (even if they are considerably shorter than the mean free path in wide
2DEG regions). Quantum wires are extremely sensitive to disorder, since the effective
scattering cross-section, being of the order of the Fermi wavelength, is comparable to
the width of the wire. Indeed, calculations demonstrate\cite{ref45} that a quantum wire close
to pinch-off breaks up into a number of isolated segments. The Coulomb-blockade oscillations in a quantum wire discussed in Sec.\ \ref{sec3} are associated with tunneling through the
barriers separating these segments (see Fig.\ \ref{fig2}a).

A quantum dot is formed in a 2DEG if the electrons are confined in all three directions. The energy spectrum of a quantum dot is fully discrete. Transport through
the discrete states in a quantum dot can be studied if tunnel barriers are defined at
its perimeter. The quantum dots discussed in Sec.\ \ref{sec3} are connected by quantum point
contacts to their surroundings (see Figs.\ \ref{fig2}b and \ref{fig2}c). The quantum point contacts are operated close to pinch-off ($G<2e^{2}/h$), where they behave as tunnel barriers of adjustable
height and width. The shape of such barriers differs greatly from that encountered in
metallic tunnel junctions: the barrier height typically exceeds the Fermi energy by only
a few meV, and the thickness of the barrier at $E_{\mathrm{F}}$ is large, on the order of 50 nm. This
may lead to a strong energy dependence of the tunnel rates, not encountered in metals.

\section{\label{sec2} Theory of Coulomb-blockade oscillations}

Part of the interest in quantum dots derives from the fact that their electronic
structure mimicks that of an isolated atom --- with the fascinating possibility to attach
wires to this ``atom'' and study transport through its discrete electronic states. In this
section we address this problem from a theoretical point of view, following Ref.\ \cite{ref19}. 

\subsection{\label{sec2.1} Periodicity of the oscillations}

We consider a quantum dot, which is weakly coupled by tunnel barriers to two electron reservoirs. A current $I$ can be passed through the dot by applying a voltage difference $V$ between the reservoirs. The linear response conductance $G$ of the quantum dot is defined as $G\equiv I/V$, in the limit $V\rightarrow 0$. Since transport through a quantum dot proceeds by tunneling through its discrete electronic states, it will be clear that for small $V$ a net current can flow only for certain vlaues of the gate voltage (if $\Delta E\gg k_{\rm B}T$). In the absence of charging effects, a conductance peak due to resonant tunneling occurs when the Fermi energy $E_{\rm F}$ in the reservoirs lines up with one of the energy levels in the dot. This condition is modified by the charging energy. To determine the location of the conductance peaks as a function of gate voltage requires only consideration of the equilibrium properties of the system,\cite{ref19,ref30} as we now discuss.

The probability $P(N)$ to find $N$ electrons in the quantum dot in equilibrium with the reservoirs is given by the grand canonical distribution function
\be
P(N)={\rm constant}\times\exp\left(-\frac{1}{k_{\rm B}T}[F(N)-NE_{\rm F}]\right), \label{eq7}
\ee
where $F(N)$ is the free energy of the dot and $T$ the temperature. The reservoir Fermi energy $E_{\rm F}$ is measured relative to the conduction band bottom in the reservoirs. In general, $P(N)$ at $T=0$ is non-zero for a {\em single\/} value of $N$ only (namely the integer which mimimizes the thermodynamic potential $\Omega(N)\equiv F(N)-NE_{\rm F}$). In that case, $G\rightarrow 0$ in the limit $T\rightarrow 0$. As pointed out by Glazman and Shekhter,\cite{ref5} a non-zero $G$ is possible only if $P(N)$ and $P(N+1)$ are both non-zero for some $N$. Then a small applied voltage is sufficient to induce a current through the dot,via intermediate states $N\rightarrow N+1\rightarrow N\rightarrow N+1\rightarrow \cdots$. To have $P(N)$ and $P(N+1)$ both non-zero at $T=0$ requires that both $N$ and $N+1$ minimize $\Omega$. A necessary condition is $\Omega(N+1)=\Omega(N)$, or
\be
F(N+1)-F(N)=E_{\rm F}.\label{eq8}
\ee
This condition is also sufficient, unless $\Omega$ has more than one minimum (which is usually not the case).

Equation (\ref{eq8}) expresses the equality of the electrochemical potential of dots and leads. The usefulness of this result is that it maps the problem of determining the location of the conductance peaks onto the more familiar problem of calculating the electrochemical potential $F(N+1)-F(N)$ of the quantum dot, i.e.\ the energy cost associated with the addition of a single electron to the dot. This opens the way, in principle, to a study of exchange and correlation effets on the conductance oscillations in a quantum dot (e.g.\ along the lines of work by Bryant\cite{ref46} and by Maksym and Chakraborty\cite{ref47}).

At $T=0$ the free energy $F(N)$ equals the ground state energy of the dot, for which we take the simplified form $U(N)+\sum_{p=1}^{N}E_{p}$. Here $U(N)$ is the charging energy, and $E_{p}$ ($p=1,2,\ldots$) are single-electron energy levels in ascending order. The term $U(N)$ accounts for the charge imbalance between dot and reservoirs. The sum over energy levels accounts for the internal degrees of freedom of the quantum dot, evaluated in a mean-field approximation (cf.\ Ref.\ \cite{ref48}). Each level contains either one or zero electrons. Spin degeneracy, if
present, can be included by counting each level twice, and other degeneracies can be
included similarly. The energy levels $E_{p}$ depend on gate voltage and magnetic field, but
are assumed to be independent of $N$, at least for the relevant range of values of $N$. We
conclude from Eq.\ (\ref{eq8}) that a peak in the low-temperature conductance occurs whenever
\be
E_{N}+U(N)-U(N-1)=E_{\mathrm{F}},   \label{eq9}
\ee
for some integer $N$ (we have relabeled $N$ by $N-1$).

We adopt the simple approximation of the orthodox model\cite{ref4} of taking the charging
energy into account macroscopically. We write $U(N)=\int_{0}^{-Ne}\phi(Q^{\prime})\mathrm{d}Q^{\prime}$, where
\be
\phi(Q)=Q/C+\phi_{\rm{ext}}   \label{eq10}
\ee
is the potential difference between dot and reservoir, including also a contribution $\phi_{\rm{ext}}$
from external charges (in particular those on a nearby gate electrode). The capacitance
$C$ is assumed to be independent of $N$ (at least over some interval). The charging energy
then takes the form
\be
U(N)=(Ne)^{2}/2C-Ne\phi_{\rm{ext}}.   \label{eq11}
\ee
To make connection with some of the literature\cite{ref3,ref49} we mention that $Q_{\rm{ext}}\equiv C\phi_{\rm{ext}}$ plays
the role of an ``externally induced charge'' on the dot, which can be varied continuously
by means of an external gate voltage (in contrast to $Q$ which is restricted to integer
multiples of $e$). In terms of $Q_{\rm{ext}}$ one can write
\[
U(N)=(Ne - Q_{\rm{ext}})^{2}/2C+ {\rm constant},
\]
which is equivalent to Eq.\ (\ref{eq11}). We emphasize that $Q_{\rm{ext}}$ is an externally controlled variable, via the gate voltage, regardless of the relative magnitude of the various capacitances
in the system.

Substitution of Eq.\ (\ref{eq11}) into Eq.\ (\ref{eq9}) gives
\be
E_{N}^{*} \equiv E_{N}+(N-\tfrac{1}{2})\frac{e^{2}}{C}=E_{\mathrm{F}}+e\phi_{\rm{ext}}   \label{eq12}
\ee
as the condition for a conductance peak. The left-hand-side of Eq.\ (\ref{eq12}) defines a renormalized energy level $E_{N}^{*}$. The renormalized level spacing $\Delta E^{*}=\Delta E+e^{2}/C$ is enhanced
above the bare level spacing by the charging energy. In the limit $e^{2}/C\Delta E\rightarrow 0$, Eq.\ (\ref{eq12})
is the usual condition for resonant tunneling. In the limit $ e^{2}/C\Delta E\rightarrow\infty$, Eq.\ (\ref{eq12}) describes the periodicity of the classical Coulomb-blockade oscillations in the conductance
versus electron density.\cite{ref3,ref4,ref5,ref6,ref7}

\begin{figure}
\centerline{\includegraphics[width=8cm]{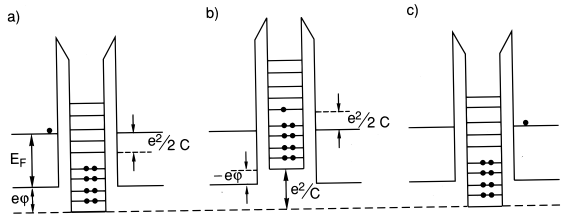}}
\caption{
Single-electron tunneling through a quantum dot, under the conditions of Eq.\ (\ref{eq12}), for
the case that the charging energy is comparable to the level spacing. An infinitesimally small voltage
difference is assumed between the left and right reservoirs. (From Beenakker et al.\cite{ref31})
\label{fig3}
}
\end{figure}

\begin{figure}
\centerline{\includegraphics[width=8cm]{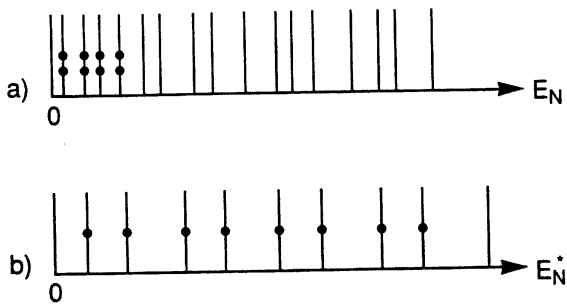}}
\caption{
Diagram of the bare energy levels (a) and the renormalized energy levels (b) in a quantum
dot for the case $e^{2}/C\approx 2\langle\Delta E\rangle$. The renormalized level spacing is much more regular than the average
bare level spacing $\langle\Delta E\rangle$. Note that the spin degeneracy of the bare levels is lifted by the charging
energy. (From Staring et al.\cite{ref12})
\label{fig4}
}
\end{figure}

In Fig.\ \ref{fig3} we have illustrated the tunneling of an electron through the dot under the
conditions of Eq.\ (\ref{eq12}). In panel (a) one has $E_{N}+e^{2}/2C=E_{\mathrm{F}}+e\phi(N-1)$, with $N$
referring to the lowest unoccupied level in the dot. In panel (b) an electron has tunneled
into the dot. One now has $E_{N}-e^{2}/2C=E_{\mathrm{F}}+e\phi(N)$, with $N$ referring to the highest
occupied level. The potential difference $\phi$ between dot and reservoir has decreased by
$e/C$ (becoming negative), because of the added electron. Finally, in panel (c) the added
electron tunnels out of the dot, resetting the potentials to the initial state of panel (a).

Let us now determine the periodicity of the oscillations. Theoretically, it is convenient to consider the case of a variation of the Fermi energy of the reservoirs at constant
$\phi_{\rm{ext}}$. The periodicity $\Delta E_{\mathrm{F}}$ follows from Eq.\ (\ref{eq12}),
\be
\Delta E_{\mathrm{F}}=\Delta E^{*}\equiv\Delta E+\frac{e^{2}}{C}.   \label{eq13}
\ee
In the absence of charging effects, $\Delta E_{\mathrm{F}}$ is determined by the irregular spacing $\Delta E$ of
the single-electron levels in the quantum dot. The charging energy $e^{2}/C$ {\it regulates\/} the
spacing, once $e^{2}/C\gtrsim \Delta E$. This is illustrated in Fig.\ 4, for the case that there is no valley
degeneracy. The spin degeneracy of the levels is lifted by the charging energy. In a
plot of $G$ versus $E_{\mathrm{F}}$ this leads to a doublet structure of the oscillations, with a spacing
alternating between $e^{2}/C$ and $\Delta E+e^{2}/C$.

\begin{figure}
\centerline{\includegraphics[width=8cm]{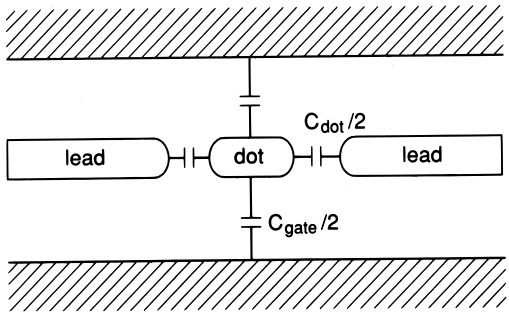}}
\caption{
Equivalent circuit of quantum dot and split gate. The mutual capacitance of leads and gate
is much larger than that of the dot and the split gate $(C_{\mathrm{gate}})$, or the dot and the leads $(C_{\mathrm{dot}})$, and can be neglected.
\label{fig5}
}
\end{figure}

Experimentally, one studies the Coulomb-blockade oscillations as a function of gate
voltage. To determine the periodicity in that case, we first need to know how $E_{\mathrm{F}}$ and
the set of energy levels $E_{p}$ depend on $\phi_{\rm{ext}}$. In a 2DEG, the external charges are supplied
by ionized donors and by a gate electrode (with an electrostatic potential difference $\phi_{\mathrm{gate}}$
between gate and 2DEG reservoir). One has
\be
\phi_{\rm{ext}}=\phi_{\rm donors}+\alpha\phi_{\rm gate}, \label{eq14}
\ee
where $\alpha$ (as well as $C$) is a rational function of the capacitance matrix elements of the
system. The value of $\alpha$ depends on the geometry. Here we consider only the geometry of
Figs.\ \ref{fig2}a,b in detail, for which it is reasonable to assume that the electron gas densities
in the dot and in the leads increase, on average, equally fast with $\phi_{\mathrm{gate}}$. For equidistant
energy levels in the dot we may then assume that $E_{\mathrm{F}}-E_{N}$ has the same value at each
conductance peak. The period of the oscillations now follows from Eqs.\ (\ref{eq12}) and (\ref{eq14}),
\be
\Delta\phi_{\mathrm{gate}}=\frac{e}{\alpha C}.   \label{eq15}
\ee
To clarify the meaning of the parameters $C$ and $\alpha$, we represent the system of dot, gates
and leads in Figs.\ \ref{fig2}a,b by the equivalent circuit of Fig.\ \ref{fig5}. The mutual capacitance of
gates and leads does not enter our problem explicitly, since it is much larger than the
mutual capacitances of gate and dot $(C_{\rm gate})$ and dot and leads $(C_{\rm dot})$. The capacitance
$C$ determining the charging energy $e^{2}/C$ is formed by $C_{\mathrm{gate}}$ and $C_{\mathrm{dot}}$ in parallel,
\be
C=C_{\mathrm{gate}}+C_{\mathrm{dot}}.   \label{eq16}
\ee
The period of the oscillations corresponds in our approximation of equidistant energy
levels ( $E_{\mathrm{F}}-E_{\mathrm{N}}={\rm constant}$) to the increment by $e$ of the charge on the dot with no
change in the voltage across $C_{\mathrm{dot}}$. This implies $\Delta\phi_{\mathrm{gate}}=e/C_{\mathrm{gate}}$, or
\be
\alpha=C_{\rm gate}/(C_{\mathrm{gate}}+C_{\mathrm{dot}}).   \label{eq17}
\ee
Thus, in terms of the electrostatic potential difference between gate and 2DEG reservoirs, the period of the conductance oscillations is $\Delta\phi_{\mathrm{gate}}=e/C_{\mathrm{gate}}$. Note that this
result applies regardless of the relative magnitudes of the bare level spacing $\Delta E$ and the
charging energy $e^{2}/C$.

In an experiment the gate voltage is the {\it electrochemical\/} potential difference $V_{\mathrm{gate}}$
between gate and leads, i.e.\ the difference in Fermi level, whereas so far we have discussed
the period of the oscillations in terms of the {\it electrostatic\/} potential difference $\phi_{\mathrm{gate}}$, i.e.\
the difference in conduction band bottoms. In one period, the change in Fermi energy
in the dot and leads (measured with respect to their local conduction band bottom) is
approximately equal to $\Delta E$. The change in Fermi energy in the (metal) gate is negligible,
because the density of states in a metal is much larger than in a 2DEG. We thus find
that the oscillation period $\Delta V_{{\rm gate}}$ in the geometry of Figs.\ \ref{fig2}a,b is
\be
\Delta V_{{\rm gate}}=\frac{\Delta E}{e}+\Delta\phi_{{\rm gate}}=\frac{\Delta E}{e}+\frac{e}{C_{{\rm gate}}}.   \label{eq18}
\ee
Note that $C_{\rm dot}$ does not affect the periodicity. In many of the present experiments $\Delta E$
is a factor of 10 below $e^{2}/C_{{\rm gate}}$, so that the differences between $\Delta\phi_{{\rm gate}}$ and $\Delta V_{{\rm gate}}$ are
less than 10 \%. Even in such a case, these differences are quite important, since their
study yields direct information on the energy spectrum of the quantum dot.

In the case of a two-fold spin-degeneracy, the level separation $E_{p+1}-E_{p}$ in a dot
of area $A$ alternates between 0 and $\Delta E\sim 2\pi\hbar^{2}/mA$ [cf.\ Eq.\ (\ref{eq3})]. As mentioned above,
this leads to a doublet structure of the oscillations as a function of $E_{\mathrm{F}}$. To determine
the peak spacing as a function of gate voltage we approximate the change in $E_{\mathrm{F}}$ with
$\phi_{{\rm gate}}$ by $\partial E_{\mathrm{F}}/\partial\phi_{{\rm gate}}\sim\Delta EC_{{\rm gate}}/2e$. We then obtain from Eqs.\ (\ref{eq12}), (\ref{eq14}), (\ref{eq16}), and (\ref{eq17})
that the spacing alternates between two values:
\begin{eqnarray}
\Delta\phi_{{\rm gate}}^{(1)}&=&\left(\frac{e}{C_{{\rm gate}}}\right)\frac{e^{2}/C}{\Delta E/2+e^{2}/C},   \label{eq19}\\
\Delta\phi_{{\rm gate}}^{(2)}&=&\left(\frac{e}{C_{{\rm gate}}}\right)\frac{\Delta E+e^{2}/C}{\Delta E/2+e^{2}/C}.   \label{eq20}
\end{eqnarray}
The average spacing equals $e/C_{{\rm gate}}$, in agreement with Eq.\ (\ref{eq15}) [derived for non-degenerate equidistant levels]. To obtain $\Delta V_{{\rm gate}}$ one has to add $\Delta E/2e$ to the factor $e/C_{{\rm gate}}$
between brackets in Eqs.\ (\ref{eq19}) and (\ref{eq20}). If the charging energy dominates ($e^{2}/C\gg\Delta E$), one has equal spacing $\Delta\phi_{\rm gate}^{(1)}=\Delta\phi_{\rm gate}^{(2)}=e/C_{\rm gate}$, as for non-degenerate levels. In the opposite limit $\Delta E\gg e^{2}/C$, one finds instead $\Delta\phi_{\rm gate}^{(1)}=0$, and $\Delta\phi_{\rm gate}^{(2)}=2e/C_{\rm gate}$. Thus, the period is effectively doubled, corresponding to the addition of {\it two\/} electrons to the dot,
instead of one. This is characteristic for resonant tunneling of non-interacting electrons
through two-fold spin-degenerate energy levels. An external magnetic field will resolve
the spin-degeneracy, leading to a splitting of the conductance peaks which increases with
the field.

\subsection{\label{sec2.2} Amplitude and lineshape}

Equation (\ref{eq12}) is sufficient to determine the periodicity of the conductance oscillations, but gives no information on their amplitude and width, which requires the solution
of a kinetic equation. For the linear response conductance in the resonant tunneling
regime an analytical solution has been derived by Beenakker,\cite{ref19} which generalizes earlier results by Kulik and Shekhter\cite{ref7} in the classical regime. Equivalent results have
been obtained independently by Meir, Wingreen, and Lee.\cite{ref20} Related work on the
non-linear current-voltage characteristics has been performed by Averin, Korotkov, and
Likharev,\cite{ref34} and by Groshev.\cite{ref35} In this sub-section we summarize the main results
of Ref.\ \cite{ref19}, along with the underlying assumptions.

A continuum of states is assumed in the reservoirs, which are occupied according
to the Fermi-Dirac distribution (\ref{eq4}). The tunnel rate from level $p$ to the left and right
reservoirs is denoted by $\Gamma_{p}^{\rm l}$ and $\Gamma_{p}^{\mathrm{r}}$, respectively. We assume that $k_{\mathrm{B}}T\gg h(\Gamma^{\rm l}+\Gamma^{\mathrm{r}})$ (for
all levels participating in the conduction), so that the finite width $h\Gamma=h(\Gamma^{\rm l}+\Gamma^{\mathrm{r}})$ of the transmission resonance through the quantum dot can be disregarded. This assumption
allows us to characterize the state of the quantum dot by a set of occupation numbers,
one for each energy level. (As we will discuss, in the classical regime $k_{\mathrm{B}}T\gg\Delta E$
the condition $\Delta E\gg h\Gamma$ takes over from the condition $ k_{\mathrm{B}}T\gg h\Gamma$ appropriate for
the resonant tunneling regime.) We assume here that inelastic scattering takes place
exclusively in the reservoirs --- not in the quantum dot. (The effects of inelastic scattering
in the dot for $ k_{\mathrm{B}}T\gg h\Gamma$ are discussed in Ref.\ \cite{ref19}.)

The equilibrium distribution function of electrons among the energy levels is given
by the Gibbs distribution in the grand canonical ensemble:
\be
P_{\mathrm{eq}}( \{n_{i}\})=\frac{1}{Z}\exp\left[-\frac{1}{k_{\mathrm{B}}T}\left(\sum_{i=1}^{\infty}E_{i}n_{i}+U(N)-NE_{\mathrm{F}}\right)\right],   \label{eq21}
\ee
where $\{n_{i}\}\equiv\{n_{1},$ $n_{2},$ $\ldots\}$ denotes a specific set of occupation numbers of the energy
levels in the quantum dot. (The numbers $n_{i}$ can take on only the values 0 and 1.) The
number of electrons in the dot is $N\equiv\sum_{i}n_{i}$, and $Z$ is the partition function,
\be
Z=\sum_{\{n_{i}\}}\exp\left[-\frac{1}{k_{\mathrm{B}}T}\left(\sum_{i=1}^{\infty}E_{i}n_{i}+U(N)-NE_{\mathrm{F}}\right)\right].   \label{eq22}
\ee
The joint probability $P_{\mathrm{eq}}(N, n_{p}=1)$ that the quantum dot contains $N$ electrons {\it and\/}
that level $p$ is occupied is
\be
P_{\mathrm{eq}}(N,n_{p}=1)= \sum_{\{n_{i}\}}P_{\mathrm{eq}}(\{n_{i}\})\delta_{N,\sum_{i}n_{i}}\delta_{n_{p},1}.   \label{eq23}
\ee
In terms of this probability distribution, the conductance is given by
\begin{eqnarray}
G&=& \frac{e^{2}}{k_{\mathrm{B}}T}\sum_{p=1}^{\infty}\sum_{N=1}^{\infty}\frac{\Gamma_{p}^{\rm l}\Gamma_{p}^{\mathrm{r}}}{\Gamma_{p}^{\rm l}+\Gamma_{p}^{\mathrm{r}}}P_{\mathrm{eq}}(N,n_{p}=1)\nonumber\\
&&\mbox{}
\times[1-f(E_{p}+U(N)-U(N-1)-E_{\mathrm{F}})].\nonumber\\
&&   \label{eq24}
\end{eqnarray}
This particular product of distribution functions expresses the fact that tunneling of
an electron from an initial state $p$ in the dot to a final state in the reservoir requires
an occupied initial state and empty final state. Equation (\ref{eq24}) was derived in Ref.\ \cite{ref19}
by solving the kinetic equation in linear response. This derivation is presented in the
appendix. The same formula has been obtained independently by Meir, Wingreen, and
Lee,\cite{ref20} by solving an Anderson model in the limit $ k_{\mathrm{B}}T\gg h\Gamma$.

We will now discuss some limiting cases of the general result (\ref{eq24}). We first consider
the conductance of the individual barriers and the quantum dot in the high temperature limit $k_{\mathrm{B}}T\gg e^{2}/C,$ $\Delta E$ where neither the discreteness of the energy levels nor the
charging energy are important. The conductance then does not exhibit oscillations as
a function of gate voltage. The high temperature limit is of interest for comparison
with the low temperature results, and because its measurement allows a straightforward
estimate of the tunnel rates through the barriers. The conductance of the quantum dot
in the high temperature limit is simply that of the two tunnel barriers in series
\be
G=\frac{G^{\rm l}G^{\mathrm{r}}}{G^{\rm l}+G^{\mathrm{r}}},\;\;{\rm if}\;\;\Delta E,e^{2}/C\ll k_{\mathrm{B}}T\ll E_{\mathrm{F}}. \label{eq25}
\ee
The conductances $G^{\rm l}$, $G^{\mathrm{r}}$ of the left and right tunnel barriers are given by the thermally
averaged Landauer formula
\be
G^{\mathrm{l,r}}=-\frac{e^{2}}{h}\int_{0}^{\infty}{\rm d}E\,T^{\mathrm{l,r}}(E)\frac{\mathrm{d}f}{\mathrm{d}E}. \label{eq26}
\ee
The transmission probability of a barrier $T(E)$ equals the tunnel rate $\Gamma(E)$ divided by
the attempt frequency $\nu(E)=1/h\rho(E)$,
\be
T^{\mathrm{l,r}}(E)=h\Gamma^{\mathrm{l,r}}(E)\rho(E).   \label{eq27}
\ee
If the height of the tunnel barriers is large, the energy dependence of the tunnel rates
and of the density of states $\rho$ in the dot can be ignored (as long as $k_{\mathrm{B}}T\ll E_{\mathrm{F}}$). The
conductance of each barrier from Eq.\ (\ref{eq26}) then becomes
\be
G^{\mathrm{l,r}}=(e^{2}/h)T^{\mathrm{l,r}}=e^{2}\Gamma^{\mathrm{l,r}}\rho \label{eq28}
\ee
(where $T$, $\Gamma$, and $\rho$ are evaluated at $E_{\mathrm{F}}$), and the conductance of the quantum dot from
Eq.\ (\ref{eq25}) is
\begin{eqnarray}
G&=&e^{2} \rho\frac{\Gamma^{\rm l}\Gamma^{\mathrm{r}}}{\Gamma^{\rm l}+\Gamma^{\mathrm{r}}}=\frac{e^{2}}{h}\frac{T^{\rm l}T^{\mathrm{r}}}{T^{\rm l}+T^{\mathrm{r}}}\equiv G_{\infty},\nonumber\\
&&{\rm if}\;\; \Delta E, e^{2}/C\ll k_{\mathrm{B}}T\ll E_{\mathrm{F}}. \label{eq29}
\end{eqnarray}
The conductance $G_{\infty}$ in the high temperature limit depends only on the barrier height
and width (which determine $T$), not on the area of the quantum dot (which determines
$\rho$ and $\Gamma$, but cancels in the expression for $G_{\infty}$).

The validity of the present theory is restricted to the case of negligible quantum
fluctuations in the charge on the dot.\cite{ref4} Since charge leaks out of the dot at a rate
$\Gamma^{\rm l}+\Gamma^{\mathrm{r}}$, the energy levels are sharply defined only if the resulting uncertainty in energy
$h(\Gamma^{\rm l}+\Gamma^{\mathrm{r}})\ll\Delta E$. In view of Eq.\ (\ref{eq27}), with $\rho\sim 1/\Delta E$, this requires $T^{\mathrm{l,r}}\ll 1$, or
$G^{\mathrm{l,r}}\ll e^{2}/h$. In the resonant tunneling regime of comparable $\Delta E$ and $k_{\mathrm{B}}T$, this criterion
is equivalent to the criterion $h\Gamma\ll k_{\mathrm{B}}T$ mentioned earlier. In the classical regime
$\Delta E\ll k_{\mathrm{B}}T$, the criterion $h\Gamma\ll\Delta E$ dominates. The general criterion $h\Gamma\ll\Delta E, k_{\mathrm{B}}T$
implies that the conductance of the quantum dot $G\ll e^{2}/h$.

\begin{figure}
\centerline{\includegraphics[width=8cm]{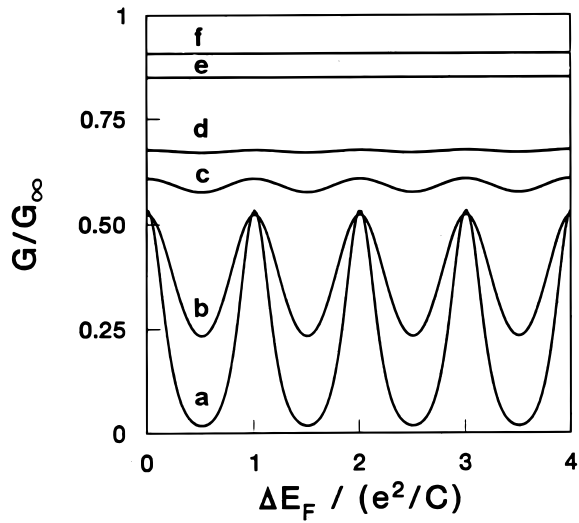}}
\caption{
Temperature dependence of the Coulomb-blockade oscillations as a function of Fermi energy
in the classical regime $k_{\mathrm{B}}T\gg\Delta E$. Curves are calculated from Eq.\ (\ref{eq24}) with $\Delta E=0.01\,e^{2}/C$, for
$k_{\mathrm{B}}T/(e^{2}/C)=$ 0.075 (a), 0.15 (b), 0.3 (c), 0.4 (d), 1 (e), and 2 (f). Level-independent tunnel rates are
assumed, as well as equidistant non-degenerate energy levels.
\label{fig6}
}
\end{figure}

\begin{figure}
\centerline{\includegraphics[width=8cm]{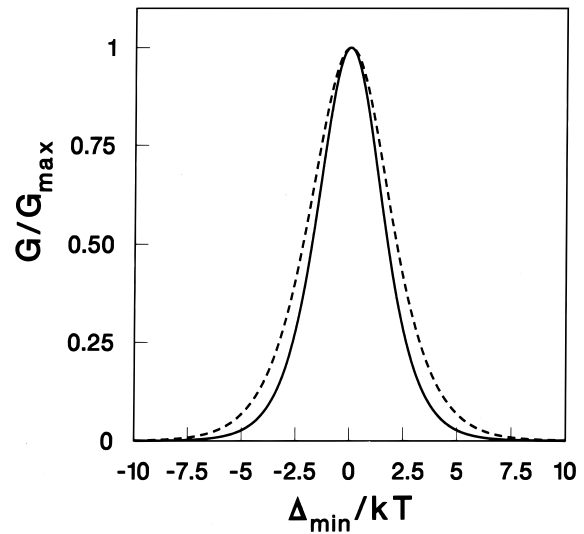}}
\caption{
Comparison of the lineshape of a thermally broadened conductance peak in the resonant tunneling regime $h\Gamma\ll k_{\rm B}T\ll\Delta E$ (solid curve) and in the classical regime $\Delta E\ll k_{\rm B}T\ll e^{2}/C$ (dashed curve). The conductance is normalized by the peak height $G_{\rm max}$, given by Eqs.\ (\ref{eq31}) and (\ref{eq34}) in the two regimes. The energy $\Delta_{\rm min}$ is proportional to the Fermi energy in the reservoirs, cf.\ Eq.\ (\ref{eq32}). (From Beenakker.\cite{ref19})
\label{fig7}
}
\end{figure}

As we lower the temperature, such that $k_{\mathrm{B}}T<e^{2}/C$, the Coulomb-blockade oscillations become observable. This is shown in Fig.\ \ref{fig6}. The classical regime $\Delta E\ll k_{\mathrm{B}}T$ was
first studied by Kulik and Shekhter.\cite{ref6,ref7} In this regime a continuum of energy levels
in the confined central region participates in the conduction. If $\Delta E\ll k_{\mathrm{B}}T\ll e^{2}/C$,
only the terms with $N=N_{\min}$ contribute to the sum in Eq.\ (\ref{eq24}), where $N_{\min}$ minimizes
the absolute value of $\Delta(N)=U(N)-U(N-1)+\overline{\mu}-E_{\mathrm{F}}$. [Here $\overline{\mu}$ is the equilibrium
chemical potential of the dot, measured relative to the bottom of the potential well.]
We define $\Delta_{\min}\equiv\Delta(N_{\min})$. For energy-independent tunnel rates and density of states
$\rho\equiv 1/\Delta E$, one obtains a line shape of individual conductance peaks given by
\begin{eqnarray}
G/G_{\max}&=& \frac{\Delta_{\min}/k_{\mathrm{B}}T}{\sinh(\Delta_{\min}/k_{\mathrm{B}}T)}\nonumber\\
&\approx&\cosh^{-2}\left(\frac{\Delta_{\min}}{2.5\,k_{\mathrm{B}}T}\right),   \label{eq30}\\
G_{\max}&=& \frac{e^{2}}{2\Delta E}\frac{\Gamma^{\rm l}\Gamma^{\mathrm{r}}}{\Gamma^{\rm l}+\Gamma^{\mathrm{r}}}.   \label{eq31}
\end{eqnarray}
The second equality in Eq.\ (\ref{eq30}) is approximate, but holds to better than 1\%. A plot of
$G/G_{\max}$ versus $\Delta_{\min}$ is shown for an isolated peak in Fig.\ \ref{fig7} (dashed curve).

Whereas the width of the peaks increases with $T$ in the classical regime, the peak
height (reached at $\Delta_{\min}=0$) is temperature independent (compare traces (a) and (b) in
Fig.\ \ref{fig6}). The reason is that the $1/T$ temperature dependence associated with resonant
tunneling through a particular energy level is canceled by the $T$ dependence of the
number $k_{\mathrm{B}}T/\Delta E$ of levels participating in the conduction. This cancellation holds only
if the tunnel rates are energy independent within the interval $k_{\mathrm{B}}T$. A temperature
dependence of the conductance may result from a strong energy dependence of the tunnel
rates. In such a case one has to use the general result (\ref{eq24}). This is also required if peaks
start to overlap for $k_{\mathrm{B}}T\sim e^{2}/C$, or if the dot is nearly depleted $(E_{\mathrm{F}}\leq k_{\mathrm{B}}T)$. The
latter regime does not play a role in metals, but is of importance in semiconductor
nanostructures because of the much smaller $E_{\mathrm{F}}$. The presence of only a small number
$E_{\mathrm{F}}/\Delta E$ of electrons in a quantum dot leads also to a gate voltage dependence of the
oscillations in the classical regime $k_{\mathrm{B}}T\gg\Delta E$.

\begin{figure}
\centerline{\includegraphics[width=8cm]{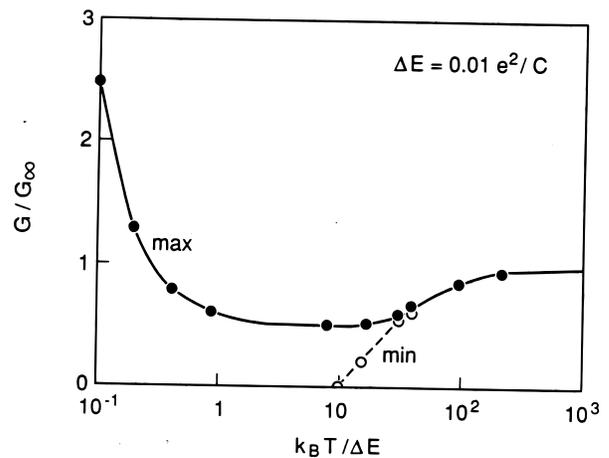}}
\caption{
Temperature dependence of the maxima (max) and the minima (min) of the Coulomb-blockade oscillations, in the regime $h\Gamma\ll k_{\rm B}T$. The calculation, based on Eq.\ (\ref{eq24}), was performed for the case of equidistant non-degenerate energy levels (at separation $\Delta E=0.01\,e^{2}/C$), all with the same tunnel rates $\Gamma^{\rm l}$ and $\Gamma^{\rm r}$.
\label{fig8}
}
\end{figure}

Despite the fact that the Coulomb blockade of tunneling is lifted at a maximum
of a conductance peak, the peak height $G_{\max}$ in the classical Coulomb-blockade regime
$\Delta E\ll k_{\mathrm{B}}T\ll e^{2}/C$ is a factor of two smaller than the conductance $G_{\infty}$ in the high
temperature regime $k_{\mathrm{B}}T\gg e^{2}/C$ of negligible charging energy (in the case of energy-independent tunnel rates). The reason is a correlation between subsequent tunnel events, imposed by the charging energy. This correlation, expressed by the series of charge states $Q=-N_{\rm min}e\rightarrow Q=-(N_{\rm min}-1)e\rightarrow Q=-N_{\rm min}e\rightarrow\cdots$, implies that an electron can tunnel from a reservoir into the dot only half of the time (when $Q=-(N_{\rm min}-1)e$). The tunnel probability is therefore reduced by a factor of two compared to the high temperature limit, where no such correlation exists.

\begin{figure}
\centerline{\includegraphics[width=8cm]{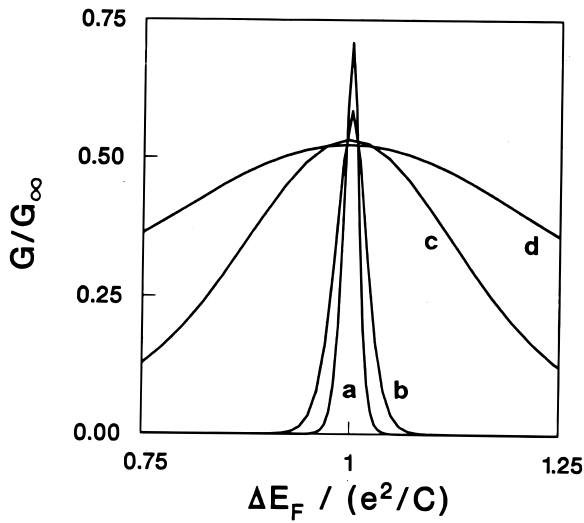}}
\caption{
Lineshape for various temperatures, showing the crossover from the resonant tunneling
regime (a and b) where both the width and the peak height depend on $T$, to the classical regime (c
and d) where only the width of the peak depends on $T$. Curves are calculated from Eq.\ (\ref{eq24}) with
$\Delta E=0.01\,\mathrm{e}^{2}/C$, and for $k_{\mathrm{B}}T/\Delta E=0.5$ (a), 1 (b), 7.5 (c), and 15 (d).
\label{fig9}
}
\end{figure}

The temperature dependence of the maxima of the Coulomb-blockade oscillations as obtained from Eq.\ (\ref{eq24}) is plotted in Fig.\ \ref{fig8}. Also shown in Fig.\ \ref{fig8} are the minima, which are seen to merge with the maxima as $k_{\rm B}T$ approaches $e^{2}/C$. In the resonant tunneling regime $k_{\rm B}T\lesssim \Delta E$ the peak height increases as the temperature is reduced, due to the diminished thermal broadening of the resonance. The crossover from the classical to the
quantum regime is shown in Fig.\ \ref{fig9} [calculated directly from Eq.\ (\ref{eq24})].

In the case of well-separated energy scales in the resonant tunneling regime $(h\Gamma\ll
k_{\mathrm{B}}T\ll\Delta E)$, Eq.\ (\ref{eq24}) can again be written in a simplified form. Now the single term
with $p=N=N_{\mathrm{min}}$ gives the dominant contribution to the sum over $p$ and $N$. The
integer $N_{\min}$ minimizes the absolute value of
\be
\Delta(N)=E_{N}+U(N)-U(N-1)-E_{\mathrm{F}}.   \label{eq32}
\ee
We again denote $\Delta_{\min}\equiv\Delta(N_{\rm min})$. Equation (\ref{eq24}) reduces to
\begin{eqnarray}
G/G_{\max}&=&4k_{\mathrm{B}}Tf'(\Delta_{\min})\nonumber\\
&=&\cosh^{-2}\left(\frac{\Delta_{\min}}{2k_{\mathrm{B}}T}\right),   \label{eq33}\\
G_{\max}&=& \frac{e^{2}}{4k_{\mathrm{B}}T}\,\frac{\Gamma^{\rm l}_{N_{\rm min}}\Gamma^{\rm r}_{N_{\rm min}}}
{\Gamma^{\rm l}_{N_{\rm min}}+\Gamma^{\rm r}_{N_{\rm min}}}.\label{eq34}
\end{eqnarray}
As shown in Fig.\ \ref{fig7}, the lineshape in the resonant tunneling regime (full curve) is different
from that in the classical regime (dashed curve), if they are compared at equal temperature. Equation (\ref{eq33}) can be seen as the usual resonant tunneling formula for a thermally
broadened resonance, generalized to include the effect of the charging energy on the resonance condition. Eqs.\ (\ref{eq33}) and (\ref{eq34})
hold regardless of the relative magnitude of $\Delta E$
and $e^{2}/C$. As illustrated in Fig.\ \ref{fig8}, the peak height in the resonant tunneling regime
increases monotonically as $k_{\mathrm{B}}T/\Delta E\rightarrow 0$, as long as $k_{\mathrm{B}}T$ is larger than the resonance width $ h\Gamma$.

No theory has been worked out for Coulomb-blockade oscillations in the regime
$k_{\mathrm{B}}T\lesssim h\Gamma$ (although the theory of Meir et al.\cite{ref20} is sufficiently general to be applicable
in principle). For {\it non-interacting\/} electrons, the transmission probability has the Breit-Wigner form\cite{ref49,ref50,ref51}
\be
G_{\mathrm{BW}}=\mathcal{G}\frac{e^{2}}{h}\frac{\Gamma^{\rm l}\Gamma^{\mathrm{r}}}{\Gamma^{\rm l}+\Gamma^{\mathrm{r}}}\frac{\Gamma}{(\epsilon/\hbar)^{2}+(\Gamma/2)^{2}}.   \label{eq35}
\ee
Here $\mathcal{G}$ is the degeneracy of the resonant level, and $\epsilon$ is the energy separation of that level
from the Fermi level in the reservoirs. In the presence of inelastic scattering with rate $\Gamma_{\mathrm{in}}$
one has to replace $\Gamma$ by $\Gamma+\Gamma_{\mathrm{in}}$.\cite{ref49,ref50,ref51} This has the effect of reducing the conductance
on resonance by a factor $\Gamma/(\Gamma+\Gamma_{\mathrm{in}})$, and to increase the width of the peak by a factor
$(\Gamma+\Gamma_{\mathrm{in}})/\Gamma$. This is to be contrasted with the regime $h\Gamma\ll k_{\mathrm{B}}T\ll\Delta E$, where inelastic
scattering has no effect on the conductance. [This follows from the fact that the thermal
average $-\int G_{\mathrm{BW}}f'(\epsilon)\mathrm{d}\epsilon\approx\int G_{\mathrm{BW}}\mathrm{d}\epsilon/4kT$ is independent of $\Gamma_{\mathrm{in}}$.] If inelastic scattering
is negligible, and if the two tunnel barriers are equal, then the maximum conductance
following from the Breit-Wigner formula is $\mathcal{G}e^{2}/h$ --- a result that may be interpreted as
the fundamental contact conductance of a $\mathcal{G}$-fold degenerate state.\cite{ref50,ref52} We surmise
that the charging energy will lift the level degeneracy, so that the maximum peak height
of Coulomb-blockade oscillations is $G_{\max}=e^{2}/h$ for the case of equal tunnel barriers.

A few words on terminology, to make contact with the resonant tunneling literature.\cite{ref49,ref50}
The results discussed above pertain to the regime $\Gamma\gg\Gamma_{\mathrm{in}}$, referred to as the
``coherent resonant tunneling'' regime. In the regime $\Gamma\ll\Gamma_{\mathrm{in}}$ it is known as ``coherent
sequential tunneling'' (results for this regime are given in Ref.\ \cite{ref19}). Phase coherence
plays a role in both these regimes, by establishing the discrete energy spectrum in the
quantum dot. The classical, or incoherent, regime is entered when $k_{\mathrm{B}}T$ or $h\Gamma_{\mathrm{in}}$ become
greater than $\Delta E$. The discreteness of the energy spectrum can then be ignored.

We close this overview of theoretical results by a discussion of the activation energy
of the minima of the conductance oscillations. It is shown in Ref.\ \cite{ref19} that $G_{\min}$ depends
exponentially on the temperature, $ G_{\min}\propto\exp(-E_{\mathrm{act}}/k_{\mathrm{B}}T)$, with activation energy
\be
E_{\mathrm{act}}=\tfrac{1}{2}(\Delta E+e^{2}/C)=\tfrac{1}{2}\Delta E^{*}.   \label{eq36}
\ee
This result holds for equal tunnel rates at two subsequent energy levels. The renormalized
level spacing $\Delta E^{*}\equiv\Delta E+e^{2}/C$, which according to Eq.\ (\ref{eq13}) determines the periodicity
of the Coulomb-blockade oscillations as a function of Fermi energy, thus equals twice the
activation energy of the conductance minima. The exponential decay of the conductance
at the minima of the Coulomb blockade oscillations results from the suppression of
tunneling processes which conserve energy in the intermediate state in the quantum dot.
Tunneling via a {\it virtual\/} intermediate state is not suppressed at low temperatures, and
may modify the temperature dependence of the minima if $ h\Gamma$ is not much smaller than
$k_{\mathrm{B}}T$ and $\Delta E$.\cite{ref53,ref54} For $h\Gamma\ll k_{\mathrm{B}}T, \Delta E$ this co-tunneling or ``macroscopic quantum tunneling of the charge'' can be neglected.

\section{\label{sec3} Experiments on Coulomb-blockade oscillations}

\subsection{\label{sec3.1} Quantum dots}

\begin{figure}
\centerline{\includegraphics[width=8cm]{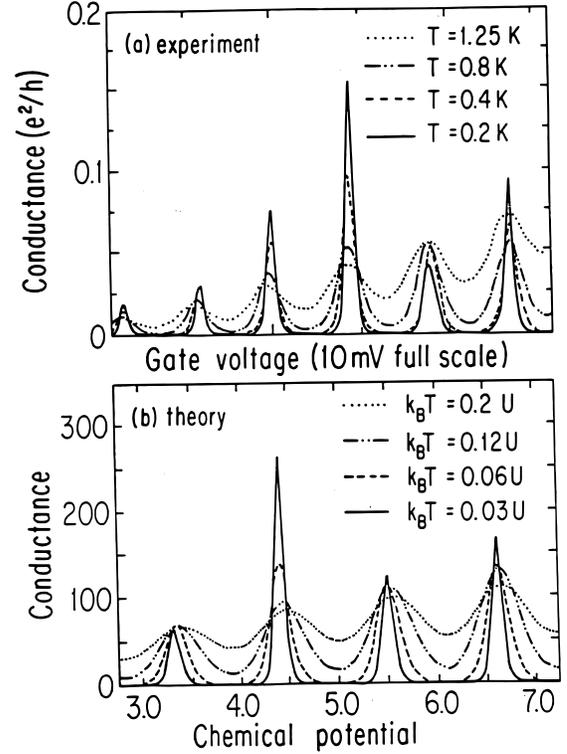}}
\caption{
(a) Measured conductance as a function of gate voltage in a quantum dot in the 2DEG of
a GaAs-AlGaAs heterostructure, with a geometry as shown in Fig.\ \ref{fig2}b. (Experimental results obtained
by U. Meirav, M. Kastner, and S. Wind, unpublished; U. Meirav, Ph.D. Thesis (M.I.T., 1990).) (b)
Calculated conductance based on Eq.\ (\ref{eq24}). The conductance is given in units $\Gamma_{1}C$, and the chemical
potential of the reservoirs in units of $e^{2}/C$. The level spacing was taken to be $\Delta E=0.1\,e^{2}/C$. The
tunnel rates of the levels increase in a geometric progression $\Gamma_{p+1}=1.5^{p}\Gamma_{1}$, with $\Gamma_{4}$ increased by an additional factor of 4 to simulate disorder. The temperature is quoted in units of $e^{2}/C$. (From Meir et
al.\ \cite{ref20}.)
\label{fig10}
}
\end{figure}

Coulomb-blockade oscillations in the conductance of a quantum dot were first studied
by Meirav, Kastner, and Wind.\cite{ref15} The geometry of their device is shown in Fig.\ \ref{fig2}b.
A split-gate electrode with a 300 nm wide slit is used to define a narrow channel. Small
protrusions on each part of the split gate are used to define quantum point contacts in the
narrow channel, 1 $\mu \mathrm{m}$ apart. For sufficiently strong negative gate voltages the electron
gas in the point contacts is depleted so that the channel is partitioned into a quantum
dot, two tunnel barriers, and two leads. The width of the quantum dot is estimated to
be 50 nm, whereas its length is about 1 $\mu \mathrm{m}$. The conductance of this device exhibits
conductance peaks periodic in the gate voltage, at temperatures between 50 mK and 1
K (see Fig.\ \ref{fig10}a). Based on estimates of the gate capacitance, it was concluded that one
electron was added to the quantum dot in each oscillation period. This conclusion was
supported by experiments on devices with different values for the tunnel barrier separation.\cite{ref15} Meirav et al.\ have also shown that the lineshape of an isolated peak could
be fitted very well by a function of the form $\cosh^{-2}(\gamma(V_{{\rm gate}}-V_{0})/2k_{\mathrm{B}}T)$. We note that,
since the fit was done with $\gamma$ and $T$ as adaptable parameters, equally good agreement
would have been obtained with the theoretical line shapes for the Coulomb-blockade
oscillations in the classical or quantum regimes [Eqs.\ (\ref{eq31}) and (\ref{eq34})].

Meirav et al. found that the temperature dependence of the peak width yielded
an estimate for $e^{2}/2C$ that was a factor of 3.5 lower than the value inferred from the
periodicity. One way to possibly resolve this discrepancy is to note that the width of the
peaks, as well as the activation energy, is determined by the charging energy $e^{2}/2C$ with
$C=C_{\rm dot}+C_{{\rm gate}}$ [Eq.\ (\ref{eq16})]. This energy is smaller than the energy $e^{2}/2C_{{\rm gate}}$ obtained
from a measurement of the periodicity $\Delta V_{{\rm gate}}\simeq e/C_{{\rm gate}}$ [Eq.\ (\ref{eq18})]. Alternatively, a
strong energy dependence of the tunnel rates may play a role.\cite{ref20}

Meir, Wingreen, and Lee\cite{ref20} modeled the experimental data shown in Fig.\ \ref{fig10}a by
means of Eq.\ (\ref{eq24}) (derived independently by these authors), using parameters consistent with experimental estimates $(\Delta E=0.1\,\mathrm{m}\mathrm{e}\mathrm{V}$, $e^{2}/C=1\,\mathrm{m}\mathrm{e}\mathrm{V})$. The results of their
calculation are reproduced in Fig.\ \ref{fig10}b. The increasing height of successive peaks is due
to an assumed increase in tunnel rates for successive levels $(\Gamma_{p+1}=1.5^{p}\Gamma_{1})$. Disorder
is simulated by multiplying $\Gamma_{4}$ by an additional factor of 4. No attempt was made to
model the gate-voltage dependence of the experiment, and instead the chemical potential of the reservoirs was chosen as a variable in the calculations. Figs.\ \ref{fig10}a and \ref{fig10}b
show a considerable similarity between experiment and theory. The second peak in the
theoretical trace is the anomalously large $\Gamma_{4}$ peak, which mimicks the fourth peak in the
experimental trace. In both theory and experiment a peak adjacent to the anomalously
large peak shows a non-monotonic temperature dependence. This qualitative agreement,
obtained with a consistent set of parameter values, supports the interpretation of the
effect as Coulomb-blockade oscillations in the regime of a discrete energy spectrum.

It is possible that at the lowest experimental temperatures in the original experiment of Meirav et al.\cite{ref15} the regime $ k_{\mathrm{B}}T\lesssim h\Gamma$ of intrinsically broadened resonances
is entered. An estimate of the average tunnel rates is most reliably obtained from the
high-temperature limit, where the peaks begin to overlap. From Fig.\ \ref{fig10}a we estimate
$G_{\infty}\sim 0.1\,e^{2}/h$. For a symmetric quantum dot $(\Gamma^{\rm l}=\Gamma^{\mathrm{r}})$ Eq.\ (\ref{eq29}) with $\rho\sim 1/\Delta E$ then
implies $h\Gamma\equiv h(\Gamma^{\rm l}+\Gamma^{\mathrm{r}})\sim 0.4\,\Delta E\sim 0.04\,\mathrm{m}\mathrm{e}\mathrm{V}$. The condition $ k_{\mathrm{B}}T\lesssim h\Gamma$ thus yields a
crossover temperature of 500 $\mathrm{m}\mathrm{K}$. Meirav et al.\cite{ref15} reported a saturation of the linear
temperature dependence of the width of the peaks to a much weaker dependence for
$T\lesssim 500\,\mathrm{m}\mathrm{K}$. It is thus possible that the approach of the intrinsically broadened regime
$ k_{\mathrm{B}}T\lesssim h\Gamma$ is at the origin of the saturated width at low temperatures (current heating of
the electron gas\cite{ref15} may also play a role). Unfortunately, as noted in Sec.\ \ref{sec2}, a theory
for the lineshape in this regime is not available.

We close the discussion of the experiments of Meirav et al.\ by noting that some
of their samples showed additional periodicities in the conductance, presumably due to
residual disorder. Thermal cycling of the sample (to room temperature) strongly affected
the additional structure, without changing the dominant oscillations due to the quantum
dot between the point contact barriers.

Williamson et al.\cite{ref17} have studied the Coulomb-blockade oscillations using a quantum dot of the design shown in Fig.\ \ref{fig2}c. The device has three sets of gates to adjust
the transmission probability of each tunnel barrier and the potential $\phi_{\rm{ext}}$ of the dot.
(Because of the proximity of the gates the adjustments are not independent.) The tunnel
barriers are formed by quantum point contacts close to pinch-off. A device with
multiple gates in a lay-out similar to that of Fig.\ \ref{fig2}b was studied by Kouwenhoven et
al.\cite{ref16} From a measurement of the Coulomb-blockade oscillations for a series of values
of the conductance of the individual quantum point contacts it has been found in both
experiments that the oscillations disappear when the conductance of each point contact
approaches the first quantized plateau, where $G^{\mathrm{l,r}}=2e^{2}/h$. It is not yet clear whether
this is due to virtual tunneling processes, or to a crossover from tunneling to ballistic
transport through the quantum point contacts. We note that this ambiguity does not
arise in tunnel junctions between metals, where the area of the tunnel barrier is usually
much larger than the Fermi wavelength squared, so that a barrier conductance larger
than $e^{2}/h$ can easily be realized {\it within\/} the tunneling regime. In semiconductors, tunnel
barriers of large area can also be made --- but it is likely that then $e^{2}/C$ will become
too small. A dynamical treatment is required in the case of low tunnel barriers, since
the field across the barrier changes during the tunnel process.\cite{ref55} Similar dynamic polarization effects are known to play a role in large-area semiconductor tunnel junctions,
where they are related to image-force lowering of the barrier height.

\subsection{\label{sec3.2} Disordered quantum wires}

Scott-Thomas et al.\cite{ref9} found strikingly regular conductance oscillations as a function
of gate voltage (or electron gas density) in a narrow disordered channel in a Si inversion
layer. The period of these oscillations differed from device to device, and did not correlate
with the channel length. Based on estimates of the sample parameters, it was concluded
that one period corresponds to the addition of a single electron to a conductance-limiting
segment of the disordered quantum wire.

Two of us have proposed that the effect is the first manifestation of {\it Coulomb-blockade
oscillations\/} in a semiconductor nanostructure.\cite{ref25} To investigate this phenomenon further, Staring et al.\ have studied the periodic conductance oscillations in disordered quantum wires defined by a split gate in the 2DEG of a GaAs-AlGaAs heterostructure.\cite{ref12,ref13}
Other studies of the effect have been made by Field et al.\cite{ref11} in a narrow channel in a
2D hole gas in Si, by Meirav et al.\cite{ref10} in a narrow electron gas channel in an inverted
GaAs-AlGaAs heterostructure, and by De Graaf et al.\cite{ref14} in a very short split gate
channel (or point contact) in a Si inversion layer. Here we will only discuss the results
of Staring et al.\ in detail.

In a first set of samples,\cite{ref12} a delta-doping layer of Be impurities was incorporated
during growth, in order to create strongly repulsive scattering centers in the narrow
channel. (Be is an acceptor in GaAs; some compensation was also present in the narrow
Si inversion layers studied by Scott-Thomas et al.\cite{ref9}) A second set of samples\cite{ref13}
did not contain Be impurities. The mean free path in the Be-doped samples in wide
regions adjacent to the channel is 0.7 $\mu \mathrm{m}$. In the other samples it is 4 $\mu \mathrm{m}$. Close to
pinch-off the channel will break up into a few segments separated by potential barriers
formed by scattering centers. Model calculations have shown that statistical variations
in the random positions of ionized donors in the AlGaAs are sufficient to create such a
situation.\cite{ref45} Indeed, both the samples with and without Be exhibited the Coulomb-blockade oscillations.

\begin{figure}
\centerline{\includegraphics[width=8cm]{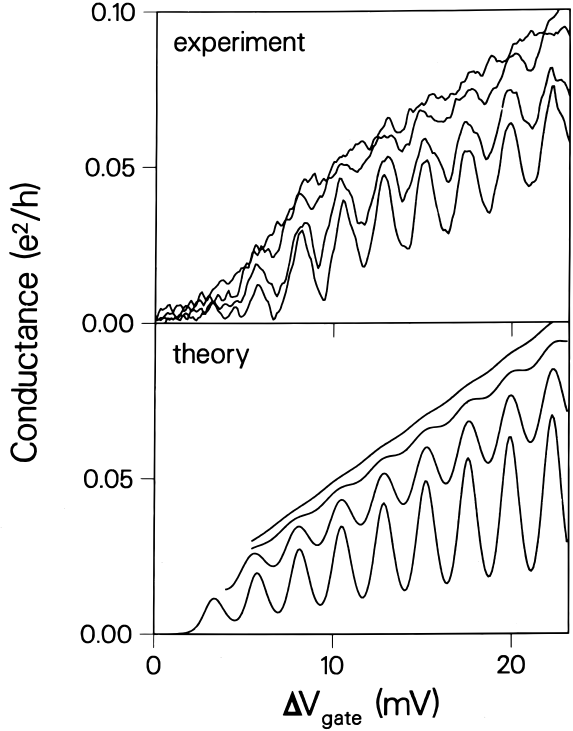}}
\caption{
(a) Measured conductance of an unintentionally disordered quantum wire in a GaAs-AlGaAs heterostructure, of a geometry as shown in Fig.\ \ref{fig2}a; $T=1.0$, 1.6, 2.5, and 3.2~K (from bottom
to top). (b) Model calculations based on Eq.\ (\ref{eq24}), for $\Delta E=0.1\,\mathrm{m}\mathrm{e}\mathrm{V}$, $e^{2}/C=0.6\,\mathrm{m}\mathrm{e}\mathrm{V}$, $\alpha=0.27$, and
$h\Gamma_{p}^{\mathrm{l,r}}=2.7\times 10^{-2}\,p\Delta E$ ($p$ labels spin-degenerate levels). (From Staring et al.\cite{ref13})
\label{fig11}
}
\end{figure}

In Fig.\ \ref{fig11}a we reproduce representative traces of conductance versus gate voltage
at various temperatures for a sample without Be.\cite{ref13} Note the similarity to the results
obtained for a single quantum dot shown in Fig.\ \ref{fig10}a. The oscillations generally disappear
as the channel is widened away from pinch-off. No correlation was found between the
periodicity of the oscillations and the channel length. At channel definition its width
equals the lithographic width $W_{\rm lith}=0.5\,\mu \mathrm{m}$, and the sheet electron density $n_{\mathrm{s}}=2.9\times
10^{11}\,\mathrm{c}\mathrm{m}^{-2}$. As the width is reduced to 0.1 $\mu \mathrm{m}$, the density becomes smaller by about a factor of 2. (The estimate for $W$ is based on typical lateral depletion widths of 200 $\mathrm{n}\mathrm{m}/\mathrm{V}$,\cite{ref8,ref45,ref46}
and that for $n_{\mathrm{s}}$ on an extrapolation of the periodicity of the Shubnikov-De
Haas oscillations.) A 3 $\mu \mathrm{m}$ long channel then contains some 450 electrons. Calculations
for a split-gate channel\cite{ref56} indicate that the number of electrons per unit length increases
approximately linearly with gate voltage. The periodicity of the conductance oscillations
as a function of gate voltage thus implies a periodicity as a function of density per unit
length.

Our model for the Coulomb-blockade oscillations in a disordered quantum wire is
essentially the same as that for a quantum dot, to the extent that a single segment
limits the conductance. To calculate $C_{\rm dot}$ and $C_{{\rm gate}}$ is a rather complicated three-dimensional electrostatic problem, hampered further by the uncertain dimensions of
the conductance limiting segment. Experimentally, the conductance peaks are spaced
by $\Delta V_{{\rm gate}}\sim 2.4\,\mathrm{m}\mathrm{V}$, so that from Eq.\ (\ref{eq18}) we estimate $C_{{\rm gate}}\sim 0.7\times 10^{-16}\,{\rm F}$. The
length $L$ of the quantum dot may be estimated from the gate voltage range $\delta V_{{\rm gate}}\sim 1\,\mathrm{V}$
between channel definition and pinch-off: $\delta V_{{\rm gate}}\sim en_{\rm s}W_{\rm lith}L/C_{{\rm gate}}$, where $n_{\mathrm{s}}$ is the
sheet density in the channel at definition. From the above estimate of $C_{{\rm gate}}$ and using
$\delta V_{{\rm gate}}\sim 1\,\mathrm{V}$, we estimate $L\sim 0.3\,\mu \mathrm{m}$.\footnote{
The estimated values for $C_{\rm gate}$ and $L$ are consistent with what one would expect for the mutual
capacitance of a length $L$ of a wire of diameter $W$ running in the middle of a gap of width $W_{\rm lith}$ in a
metallic plane (the thickness of the AlGaAs layer between the gate and the 2DEG is small compared
to $W_{\rm lith}$): $C_{{\rm gate}}\sim 4\pi\epsilon L/2\,{\rm arccosh}\,(W_{\rm lith}/W)\sim 0.9\times 10^{-16}\,\mathrm{F}$ (see Ref.\ \cite{ref57}).}
The width of the dot is estimated to be about
$W\sim 0.1\,\mu \mathrm{m}$ in the gate voltage range of interest. The level splitting in the segment
is $\Delta E\sim 2\pi\hbar^{2}/mLW\sim 0.2\,\mathrm{m}\mathrm{e}\mathrm{V}$ (for a 2-fold spin-degeneracy). Since each oscillation
corresponds to the removal of a single electron from the dot, the maximum number of
oscillations following from $\Delta E$ and the Fermi energy $E_{\mathrm{F}}\sim 5\,\mathrm{m}\mathrm{e}\mathrm{V}$ at channel definition
is given by $2E_{\mathrm{F}}/\Delta E\sim 50$, consistent with the observations. From the fact that the
oscillations are still observable at $T=1.5\,\mathrm{K}$, albeit with considerable thermal smearing,
we deduce that in our experiments $e^{2}/C+\Delta E\sim 1\,\mathrm{m}\mathrm{e}\mathrm{V}$. Thus, $C\sim 2.0\times 10^{-16}\,\mathrm{F}$,
$C_{\rm dot}=C-C_{{\rm gate}}\sim 1.3\times 10^{-16}\,\mathrm{F}$,\footnote{
The mutual capacitance of dot and leads may be approximated by the self-capacitance of the dot,\cite{ref57}
which should be comparable to that of a two-dimensional circular disc of diameter $ L$: $C_{\rm dot}\sim 4\epsilon L\sim 1.4\times 10^{-16}\,\mathrm{F}$, consistent with the estimate given in the text.}
and the parameter $\alpha\equiv C_{{\rm gate}}/C\sim 0.35$. In Fig.\ \ref{fig11}
we compare a calculation based on Eq.\ (\ref{eq24}) with the experiment, taking the two-fold
spin-degeneracy of the energy levels into account\cite{ref13} The tunnel rates were taken to
increase by an equal amount $0.027\,\Delta E/h$ for each subsequent spin-degenerate level, at
equal separation $\Delta E=0.1\,\mathrm{m}\mathrm{e}\mathrm{V}$. The capacitances were fixed at $e^{2}/C=0.6\,\mathrm{m}\mathrm{e}\mathrm{V}$ and
$\alpha=0.25$. These values are consistent with the crude estimates given above. The Fermi
energy was assumed to increase equally fast as the energy of the highest occupied level
in the dot (cf.\ Sec.\ \ref{sec2.1}). The temperature range shown in Fig.\ \ref{fig11} is in the classical
regime $(k_{\mathrm{B}}T>\Delta E)$.

\begin{figure}
\centerline{\includegraphics[width=8cm]{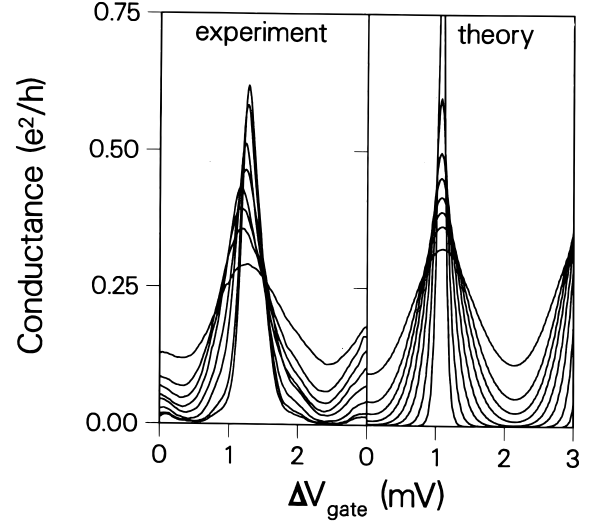}}
\caption{
Experimental and theoretical lineshapes of an isolated conductance peak in a Be-doped
disordered quantum wire in a GaAs-AlGaAs heterostructure, at $B=6.7\,\mathrm{T}$, and $T=110$, 190, 290, 380,
490, 590, 710, and 950 $\mathrm{m}\mathrm{K}$ (from top to bottom). The theoretical curves have been calculated from
Eq.\ (\ref{eq24}), with $\Delta E= 0.044\,\mathrm{m}\mathrm{e}\mathrm{V}$ (non-degenerate), $e^{2}/C=0.53\,\mathrm{m}\mathrm{e}\mathrm{V}$, $h\Gamma=0.13\,\mathrm{m}\mathrm{e}\mathrm{V}$, and $\alpha=0.27$.
(From Staring et al.\cite{ref13})
\label{fig12}
}
\end{figure}

The resonant tunneling regime $k_{\mathrm{B}}T<\Delta E$ can be described qualitatively by Eq.\ (\ref{eq24}),
as shown in Fig.\ \ref{fig12} for an isolated peak. The data was obtained for a different sample
(with Be doping) in the presence of a magnetic field of 6.7 T. The parameter values
used are $\Delta E= 0.045\,\mathrm{m}\mathrm{e}\mathrm{V}$, $e^{2}/C=0.53\,\mathrm{m}\mathrm{e}\mathrm{V}$, $h\Gamma=0.13\,\mathrm{m}\mathrm{e}\mathrm{V}$, and $\alpha=0.27$. A fully
quantitative theoretical description of the experimental lineshapes in Fig.\ \ref{fig12} is not yet
possible, because the experiment is in the regime of intrinsically broadened resonances,
$ k_{B}T<h\Gamma$, for which the theory has not been worked out.

The semi-quantitative agreement between theory and experiment in Figs.\ \ref{fig11} and \ref{fig12},
for a consistent set of parameter values, and over a wide range of temperatures, supports
our interpretation of the conductance oscillations as Coulomb-blockade oscillations in the
regime of comparable level spacing and charging energies. Note that $e^{2}/C_{{\rm gate}}\sim 10\,\Delta E$,
so that irregularly spaced energy levels would not easily be discernable in the gate voltage
scans [cf.\ Eq.\ (\ref{eq18})]. Such irregularities might nevertheless play a role in causing peak
height variations. Some of the data (not shown) exhibits beating patterns,\cite{ref12,ref13} similar
to those reported in Refs.\ \cite{ref9} and \cite{ref11}. These are probably due to the presence of multiple
segments in the quantum wires.\cite{ref13} Coulomb-blockade oscillations in arrays of tunnel
junctions in the classical regime have been studied by several authors,\cite{ref58,ref59}

As an alternative explanation of the conductance oscillations resonant tunneling
of non-interacting electrons has been proposed.\cite{ref26,ref27} There are several compelling
arguments for rejecting this explanation (which apply to the experiments on a quantum
dot as well as to those on disordered quantum wires). Firstly, for resonant tunneling the
oscillations would be irregularly spaced, due to the non-uniform distribution of the bare
energy levels [cf.\ Eq.\ (\ref{eq20})]. This is in contradiction with the experimental observations.\cite{ref11}
Secondly,\cite{ref12} in the absence of charging effects the measured activation energy
of the conductance minima would imply a level spacing $\Delta E\sim 1\,\mathrm{m}\mathrm{e}\mathrm{V}$. Since the Fermi
energy $E_{\mathrm{F}}$ in a typical narrow channel is about 5 $\mathrm{m}\mathrm{e}\mathrm{V}$, such a large level spacing would
restrict the possible total number of oscillations in a gate voltage scan to $E_{\mathrm{F}}/\Delta E\sim 5$,
considerably less than the number seen experimentally.\cite{ref9,ref12} Thirdly, one would expect
a spin-splitting of the oscillations by a strong magnetic field, which is not observed.\cite{ref11}
Finally, the facts that no oscillations are found as a function of {\it magnetic\/} field\cite{ref11,ref12}
and that the spin-splitting does not occur, all but rule out resonant tunneling of non-interacting electrons as an explanation of the oscillations as a function of gate voltage.

\subsection{\label{sec3.3} Relation to earlier work on disordered quantum wires}

\begin{figure}
\centerline{\includegraphics[width=8cm]{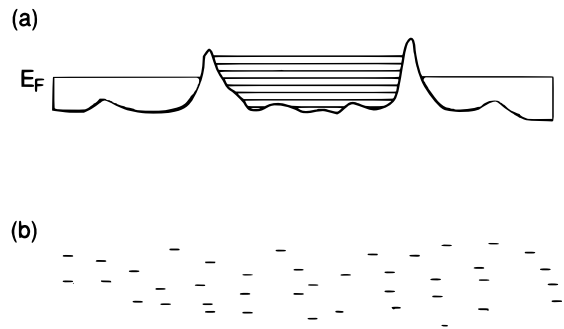}}
\caption{
(a) Coulomb-blockade oscillations occur in a disordered quantum wire as a result of the
formation of a conductance limiting segment which contains many localized states. (b) Random conductance fluctuations due to variable range hopping between localized states (indicated by dashes) are
found in the absence of such a segment.
\label{fig13}
}
\end{figure}

The disordered quantum wires discussed in this chapter exhibit {\it periodic\/} conductance
oscillations as a function of gate voltage. The effect has been seen in electron and
hole gases in Si\cite{ref9,ref11,ref14} and in the electron gas in GaAs.\cite{ref10,ref12,ref13} In contrast,
previous work by Fowler et al.\cite{ref60} and by Kwasnick et al.\cite{ref61} on narrow inversion and
accumulation layers in Si has produced sharp but {\it aperiodic\/} conductance peaks. How
are these observations to be reconciled? We surmise that the explanation is to be found
in the different strength and spatial scale of the potential fluctuations in the wire, as
illustrated in Fig.\ \ref{fig13}.

Coulomb-blockade oscillations require a small number of large potential spikes, so
that a single segment limits the conductance (Fig.\ \ref{fig13}a). The random conductance fluctuations seen previously\cite{ref60,ref61} are thought instead to be due to variable range hopping
between a large number of localized states, distributed randomly along the length of the
channel (Fig.\ \ref{fig13}b).\cite{ref62,ref63,ref64} No segment containing a large number of states (localized
within the same region) is present in the potential of Fig.\ \ref{fig13}b, in contrast to the situation
shown in Fig.\ \ref{fig13}a. At large Fermi energy a transition eventually occurs to the diffusive transport regime in either type of wire. Both the regular Coulomb-blockade oscillations,
and the random conductance peaks due to variable range hopping are then replaced by
``universal'' conductance fluctuations caused by quantum interference.\cite{ref65,ref66}

Fowler et al.\cite{ref67} have also studied the conductance of much shorter channels than
in Ref.\ \cite{ref60} (0.5 $\mu \mathrm{m}$ long, and 1 $\mu \mathrm{m}$ wide). In such channels they found well-isolated
conductance peaks, which were temperature independent below 100 $\mathrm{m}\mathrm{K}$, and which were
attributed to resonant tunneling. At very low temperatures a fine structure (some of it
time-dependent) was observed. A numerical simulation\cite{ref68} of the temporal fluctuations
in the distribution of electrons among the available sites also showed fine structure if
the time scale of the fluctuations is short compared to the measurement time, but large
compared to the tunnel time. It is possible that a similar mechanism causes the fine
structure on the Coulomb-blockade oscillations in a disordered quantum wire (cf.\ Fig.\ \ref{fig11}).

There have also been experimental studies of the effect of a strong magnetic field on
variable range hopping\cite{ref69} and on resonant tunneling through single impurity states.\cite{ref70}
We briefly discuss the work on resonant tunneling by Kopley et al.,\cite{ref70} which is more
closely related to the subject of this chapter. They observed large conductance peaks
in a Si inversion layer under a split gate. Below the 200 $\mathrm{n}\mathrm{m}$ wide slot in the gate the
inversion layer is interrupted by a potential barrier. Pronounced conductance peaks were
seen at 0.5 $\mathrm{K}$ as the gate voltage was varied in the region close to threshold. The peaks
were attributed to resonant tunneling through single impurity states in the Si bandgap
in the barrier region. The lineshape of an isolated peak could be fitted with the Breit-Wigner formula [Eq.\ (\ref{eq35})]. The amplitude of most peaks was substantially suppressed
on applying a strong magnetic field. This was interpreted as a reduction of the tunnel
rates because of a reduced overlap between the wavefunctions on the (asymmetrically
placed) impurity and the reservoirs. The amplitude of one particular peak was found to
be unaffected by the field, indicative of an impurity which is placed symmetrically in the
barrier ($\Gamma^{\mathrm{r}}=\Gamma^{\rm l}$). The width of that peak was reduced, consistent with a reduction of $\Gamma$. This study therefore exhibits many characteristic features of resonant tunneling through
a single localized site. Yet, one would expect Coulomb interactions of two electrons on
the site to be important, and indeed they might explain the absence of spin-splitting
of the peaks in a strong magnetic field.\cite{ref70} Theoretical work indicates that Coulomb
interactions also modify the lineshape of a conductance peak.\cite{ref68,ref71} The experimental
evidence\cite{ref63,ref67,ref69,ref70} is not conclusive, however.

\section{\label{sec4} Quantum Hall effect regime}

\subsection{\label{sec4.1} The Aharonov-Bohm effect in a quantum dot}

The Aharonov-Bohm effect is a quantum interference effect which results from the
influence of the vector potential on the phase of the electron wavefunction. Aharonov
and Bohm\cite{ref72} originally considered the influence of the vector potential on electrons
confined to a multiply-connected region (such as a ring), within which the magnetic field
is zero. The ground state energy of the system is periodic in the enclosed flux with
period $h/e$, as a consequence of gauge invariance. Coulomb repulsion does not affect this
periodicity.

In the solid state, the Aharonov-Bohm effect manifests itself as a periodic oscillation
in the conductance of a sample as a function of an applied magnetic field $\bm{B}$. A well-defined
periodicity requires that the conducting paths through the sample enclose a constant
area $A$, perpendicular to $\bm{B}$. The periodicity of the oscillations is then $\Delta B=h/eA$, plus
possibly harmonics (e.g.\ at $h/2eA$). The constant area may be imposed by confining the
electrons electrostatically to a ring or to a cylindrical film.\cite{ref73,ref74}

Entirely new mechanisms for the Aharonov-Bohm effect become operative in strong
magnetic fields in the quantum Hall effect regime. These mechanisms do not require a
ring geometry, but apply to singly-connected geometries such as a point contact\cite{ref75} or
a quantum dot.\cite{ref28,ref29} As discussed below, these geometries behave as if they were
multiply connected, because of circulating edge states. Resonant tunneling through
these states leads to magnetoconductance oscillations with a fundamental periodicity
$\Delta B=h/eA$, governed by the addition to the dot of a single quantum of magnetic flux
$h/e$.

An essential difference with the original Aharonov-Bohm effect is that in these experiments the magnetic field extends into the conducting region of the sample. Since
the periodicity is now no longer constrained by gauge invariance, this opens up the possibility, in principle, of an influence of Coulomb repulsion. We will discuss in the next
subsection that the Aharonov-Bohm effect may indeed be {\it suppressed\/} by charging effects.\cite{ref30}
In this subsection we will first introduce the case of negligible charging effects in
some detail.

If one applies a magnetic field $\bm{B}$ to a metal, then the electrons move with constant
velocity $v_{\parallel}$ in a direction parallel to $\bm{B}$, and in a circular cyclotron orbit with tangential
velocity $v_{\perp}$ in a plane perpendicular to $\bm{B}$. The cyclotron frequency is $\omega_{\mathrm{c}}=eB/m$, and
the cyclotron radius is $l_{\mathrm{cycl}}=v_{\perp}/\omega_{\mathrm{c}}$. Quantization of the periodic cyclotron motion in a
strong magnetic field leads to the formation of Landau levels
\begin{eqnarray}
&&E_{n}(k_{\parallel})=E_{n}+\frac{\hbar^{2}k_{\parallel}^{2}}{2m},\label{eq37}\\
&&E_{n}=(n-\tfrac{1}{2})\hbar\omega_{\mathrm{c}},\label{eq38}
\end{eqnarray}
labeled by the Landau level index $n=1,2,\ldots$. In a field of 10 $\mathrm{T}$ (which is the strongest
field that is routinely available), the Landau level separation $\hbar\omega_{\mathrm{c}}$ is about 1 $\mathrm{m}\mathrm{e}\mathrm{V}$ (for
$m=m_{\mathrm{e}})$. Consequently, in a metal the number of occupied Landau levels $N_{\mathrm{L}}\sim E_{\mathrm{F}}/\hbar\omega_{\mathrm{c}}$
is a large number, of order 1000. Even so, magnetic quantization effects are important
at low temperatures, since $\hbar\omega_{\mathrm{c}}>k_{\mathrm{B}}T$ for $T<10$ K. A familiar example is formed by
the Shubnikov-De Haas oscillations in the magnetoresistance, which are caused by peaks
in the density of states at the energies $E_{n}$ which coincide with $E_{\mathrm{F}}$ for successive values
of $n$ as $B$ is varied.

\begin{figure}
\centerline{\includegraphics[width=8cm]{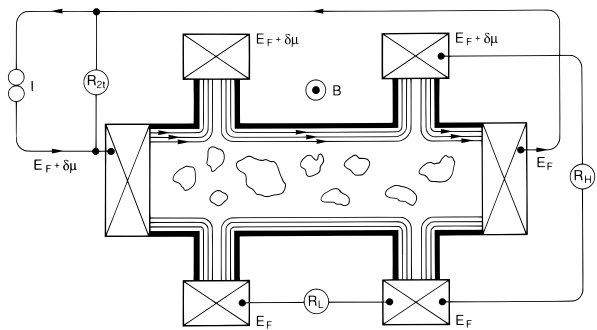}}
\caption{
Measurement configuration for the two-terminal resistance $R_{2\mathrm{t}}$, the four-terminal Hall resistance $R_{\mathrm{H}}$, and the longitudinal resistance $R_{\mathrm{L}}$. The $N_{\mathrm{L}}$ edge channels at the Fermi level are indicated,
arrows point in the direction of motion of edge channels filled by the source contact at chemical potential $E_{\mathrm{F}}+\delta\mu$. The current $N_{\mathrm{L}}e\delta\mu/h$ is equipartitioned among the edge channels at the upper edge,
corresponding to the case of local equilibrium. Localized states in the bulk do not contribute to the
conductance. The resulting resistances are $R_{2\mathrm{t}}=R_{\mathrm{H}}=h/N_{\mathrm{L}}e^{2}$, $R_{\mathrm{L}}=0$. (From Beenakker and Van
Houten.\cite{ref8})
\label{fig14}
}
\end{figure}

Because of the free motion along $\bm{B}$, the density of states in a metal does not vanish at
energies between two Landau levels. Consequently, in metals magnetic quantum effects
are relatively small. The situation is different in a 2DEG. Here the energy spectrum of
the electrons becomes fully discrete in a strong perpendicular magnetic field, since no
free translational motion parallel to $\bm{B}$ is possible. The vanishing of the density of states
between Landau levels is at the origin of the pronounced magnetic quantum effects in
a 2DEG. Well known is the integer quantum Hall effect, characterized by a vanishing
longitudinal resistance $R_{\mathrm{L}}$ and a quantized Hall resistance $R_{\mathrm{H}}$ at values of $h/N_{\mathrm{L}}e^{2}$. The
distinction between a longitudinal and Hall resistance is topological (see Fig.\ \ref{fig14}): A
four-terminal resistance measurement gives $R_{\mathrm{H}}$ if current and voltage contacts alternate
along the boundary of the conductor, and $R_{\mathrm{L}}$ if that is not the case. There is no need
to further characterize the contacts in the case of local equilibrium at the edge (in the
opposite case the Hall resistance may take on anomalous values\cite{ref8}). Frequently, the
resistance of a sample is measured using only two contacts (which then act both as
current and as voltage probes). In the quantum Hall effect regime, the two-terminal
resistance $R_{2\mathrm{t}}=R_{\mathrm{H}}+R_{\mathrm{L}}=R_{\mathrm{H}}$ is quantized at the same value as the Hall resistance.

The Fermi energy in a 2DEG is quite small (10 $\mathrm{m}\mathrm{e}\mathrm{V}$ in conventional samples, 1 $\mathrm{m}\mathrm{e}\mathrm{V}$
for samples with a very low density $n_{\mathrm{s}}\sim 10^{10}\mathrm{c}\mathrm{m}^{-2}$). Since, in addition, the effective
mass is small, the extreme magnetic quantum limit $N_{\mathrm{L}}=1$ is accessible. This is the realm
of the fractional quantum Hall effect, studied in high-mobility samples at milli-Kelvin
temperatures, and of the Wigner crystallization of the 2DEG. Both phenomena are due
to electron-electron interactions in a strong magnetic field. This chapter is limited to
the integer quantum Hall effect.

To the extent that broadening of the Landau levels by disorder can be neglected,
the density of states (per unit area) in an unbounded 2DEG can be approximated by a
series of delta functions,
\be 
\rho(E)=g_{\mathrm{s}}g_{\mathrm{v}}\frac{eB}{h}\sum_{n=1}^{\infty}\delta(E-E_{n}).\label{eq39}
\ee
The spin-degeneracy $g_{\mathrm{s}}$ is removed in strong magnetic fields as a result of the Zeeman
splitting $g\mu_{\rm B}B$ of the Landau levels ($\mu_{\rm B}\equiv e\hbar/2m_{\rm e}$ denotes the Bohr magneton; the Lande $g$-factor is a complicated function of the magnetic field in these systems\cite{ref76}).

In the modern theory of the quantum Hall effect,\cite{ref77} the longitudinal and Hall
conductance (measured using two pairs of current contacts and voltage contacts) are
expressed in terms of the transmission probabilities between the contacts for electronic
states at the Fermi level. When $E_{\mathrm{F}}$ lies between two Landau levels, these states are {\it edge states\/} extended along the boundaries (Fig.\ \ref{fig14}). Edge states are the quantum mechanical
analogue of {\it skipping orbits\/} of electrons undergoing repeated specular reflections at the
boundary.\cite{ref8} For a smooth confining potential $V(\bm{r})$, the edge states are extended along
equipotentials of $V$ at the guiding center energy $E_{\mathrm{G}}$, defined by
\be
E_{\mathrm{G}}=E-(n-\tfrac{1}{2})\hbar\omega_{\mathrm{c}},   \label{eq40}
\ee
for an electron with energy $E$ in the $n-$th Landau level $(n=1,2,\ldots)$. The confining
potential should be sufficiently smooth that it does not induce transitions between different
values of $n$. This requires that $l_{\mathrm{m}}V'\lesssim\hbar\omega_{\mathrm{c}}$, with $l_{\mathrm{m}}\equiv(\hbar/eB)^{1/2}$ the magnetic length
(which plays the role of the wave length in the quantum Hall effect regime). Since the
lowest Landau level has the largest guiding center energy, the corresponding edge state
is located closest to the boundary of the sample, whereas the higher Landau levels are
situated further towards its center.

In an open system, the single-electron levels with quantum number $n$ form a 1D
subband with subband bottom at $E_{n}=(n-\frac{1}{2})\hbar\omega_{\mathrm{c}}$. These 1D subbands are referred to
as {\it edge channels}. Each of the $N_{\mathrm{L}}\sim E_{\mathrm{F}}/\hbar\omega_{\mathrm{c}}$ edge channels at the Fermi level contributes
$2e^{2}/h$ to the Hall conductance if backscattering is suppressed. This happens whenever
the Fermi level is located between two bulk Landau levels, so that the only states at
$E_{\mathrm{F}}$ are those extended along the boundaries. Backscattering then requires transitions
between edge states on {\it opposite\/} boundaries, which are usually far apart. In a very narrow
channel, the Hall conductance may deviate from its quantized value $N_{\mathrm{L}}e^{2}/h$ (and the
longitudinal resistance may become non-zero) due to tunneling between opposite edges
--- a process that is strongly enhanced by disorder in the channel. The reason is that
localized states at the Fermi energy may act as intermediate sites in a tunneling process
from one edge to the other. We will come back to this point at the end of the section.

\begin{figure}
\centerline{\includegraphics[width=8cm]{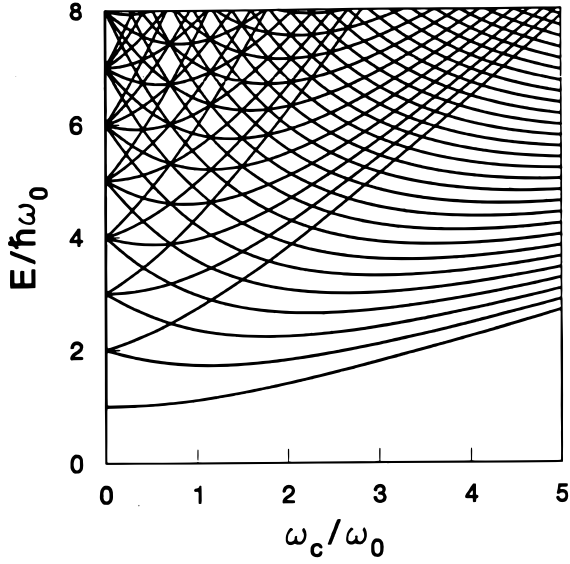}}
\caption{
Energy spectrum of a quantum dot with a harmonic confining potential as a function of
magnetic field, according to Eq.\ (\ref{eq41}). Spin-splitting is neglected.
\label{fig15}
}
\end{figure}

In a closed system, such as a quantum dot, the energy spectrum is fully discrete (for
$E_{\mathrm{G}}$ less than the height $E_{\mathrm{B}}$ of the tunnel barriers which connect the dot to the leads).
An example which can be solved exactly is a quantum dot defined by a 2D harmonic
oscillator potential $V(r)=\frac{1}{2}m\omega_{0}^{2}r^{2}$. The energy spectrum is given by\cite{ref78,ref79}
\begin{eqnarray}
E_{nm}&=&\tfrac{1}{2}(n-m)\hbar\omega_{\mathrm{c}}+\tfrac{1}{2}\hbar(\omega_{\mathrm{c}}^{2}+4\omega_{0}^{2})^{1/2}(n+m-1),\nonumber\\
&&n,m=1,2,\ldots   \label{eq41}
\end{eqnarray}
Each level has a two-fold spin-degeneracy, which is gradually lifted as $B$ is increased. For
simplicity, we do not take the spin degree of freedom into account. The energy spectrum
(\ref{eq41}) is plotted in Fig.\ \ref{fig15}. The asymptotes corresponding to the first few Landau levels
are clearly visible.

In the limit $\omega_{0}/\omega_{\mathrm{c}}\rightarrow 0$ of a smooth potential and a fairly strong magnetic field,
Eq.\ (\ref{eq41}) reduces to
\be
E_{nm}=\hbar\omega_{\mathrm{c}}(n-\tfrac{1}{2}+(n+m-1)(\omega_{0}/\omega_{\mathrm{c}})^{2}),   \label{eq42}
\ee
which may also be written as
\be
E_{nm}=(n-\tfrac{1}{2})\hbar\omega_{\rm c}+V(R_{nm}),\;\;B\pi R_{nm}^{2}=(m+\gamma_{n})\frac{h}{e},\label{eq43}
\ee
with $\gamma_{n}=n-1$. Equation (\ref{eq43}) is equivalent to the requirement that the equipotential of the edge state, of radius $R_{nm}$, encloses $m+\gamma_{n}$ flux quanta. This geometrical requirement holds generally for smooth confining potentials, in view of the Bohr-Sommerfeld quantization rule
\be
\frac{1}{h}\oint PdQ=m+\gamma_{n}.\label{eq44}
\ee
The canonically conjugate variables $P$ and $Q$, in the present case, are proportional to
the guiding center coordinates $\bm{R}=(X,Y)$, defined by
\begin{eqnarray}
X&=&x-v_{y}/\omega_{\mathrm{c}},   \label{eq45}\\
Y&=&y+v_{x}/\omega_{\mathrm{c}},   \label{eq46}
\end{eqnarray}
in terms of the position $\bm{r}=(x,y)$ and velocity $\bm{v}=(v_{x},v_{y})$ of the electron. If one
identifies $Q\equiv X$, $P\equiv eBY$, one can verify the canonical commutation relation $[Q,P]=i\hbar$ (using $m\bm{v}=\bm{p}+e\bm{A}$, $[x,p_{x}]=[y,p_{y}]=i\hbar$, $[p_{y},A_{x}]-[p_{x},A_{y}]=i\hbar B$). The Bohr-Sommerfeld quantization rule thus becomes
\be 
\Phi=B\oint YdX=\frac{h}{e}(m+\gamma_{n}),   \label{eq47}
\ee
which is the requirement that the flux $\Phi$ enclosed by the guiding center drift is quantized
in units of the flux quantum. To close the argument, we compute the guiding center drift
$\dot{\bm{R}}=B^{-2}\bm{E}(\bm{r})\times \bm{B}\sim B^{-2}\bm{E}(\bm{R})\times \bm{B}$, in the approximation that the electric field $\bm{E}$ does
not vary strongly over the cyclotron radius $|\bm{r}-\bm{R}|$. In this case of a smoothly varying $V$,
the motion of $\bm{R}$ is along equipotentials at the guiding center energy $E_{\mathrm{G}}=E-(n-\frac{1}{2})\hbar\omega_{\mathrm{c}}$.
The Bohr-Sommerfeld quantization rule can thus be written in the general form
\be
E_{nm}=(n-\tfrac{1}{2})\hbar\omega_{\mathrm{c}}+E_{\mathrm{G}}(n,m),   \label{eq48}
\ee
where $E_{\mathrm{G}}(n,m)$ is the energy of the equipotential which encloses $m+\gamma_{n}$ flux quanta. For
the harmonic oscillator potential, $\gamma_{n}=n-1$. For other smooth confining potentials $\gamma_{n}$
may be different. (Knowledge of $\gamma_{n}$ is not important if one only considers states within
a single Landau level.)

\begin{figure}
\centerline{\includegraphics[width=8cm]{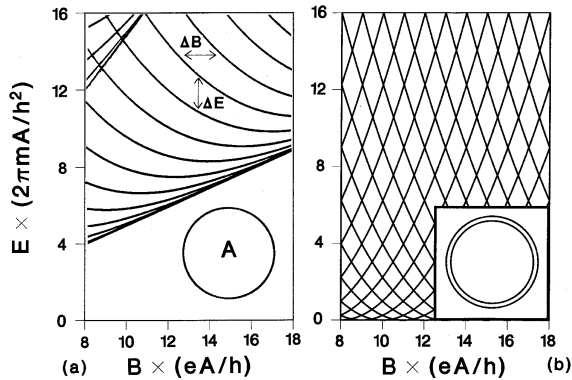}}
\caption{
Comparison of the energy levels in a disc and a ring. (a) Circular hard-wall disc (after
Geerinckx et al.\cite{ref80}). (b) Circular channel or ring of width $W\ll l_{\mathrm{m}}$ (after B\"{u}ttiker et al.\cite{ref81}). The
levels in (b) are plotted relative to the energy of the bottom of the one-dimensional subband in the
channel. The case $W\gtrsim l_{\mathrm{m}}$ is qualitatively the same as long as the area $S$ of the annulus is much smaller
than the area $A$. Spin-splitting is disregarded. (From Beenakker et al.\cite{ref30})
\label{fig16}
}
\end{figure}

Equation (\ref{eq48}) does not hold for a hard-wall confining potential. An exact solution
exists in this case for a circular disc\cite{ref80} of radius $R$, defined by $V(r)=0$ for $r<R$,
and $ V(r)=\infty$ for $r>R$. The case of a square disc was studied numerically by Sivan
et al.\cite{ref29} In Fig.\ \ref{fig16}a we show the energy spectrum as a function of $B$ for the circular
disc. (Fig.\ \ref{fig16}b is discussed in the following subsection.) The asymptotes correspond
to the bulk Landau levels $E_{n}=(n-\frac{1}{2})\hbar\omega_{\mathrm{c}}$. The first two Landau levels $(n=1,2)$
are visible in Fig.\ \ref{fig16}a. The states between the Landau levels are edge states, which
extend along the perimeter of the disc. These circulating edge states make the geometry
effectively doubly connected --- in the sense that they enclose a well-defined amount of
flux. Resonant tunneling through these states is the mechanism leading to the Aharonov-
Bohm magnetoconductance oscillations in a quantum dot.

\begin{figure}
\centerline{\includegraphics[width=8cm]{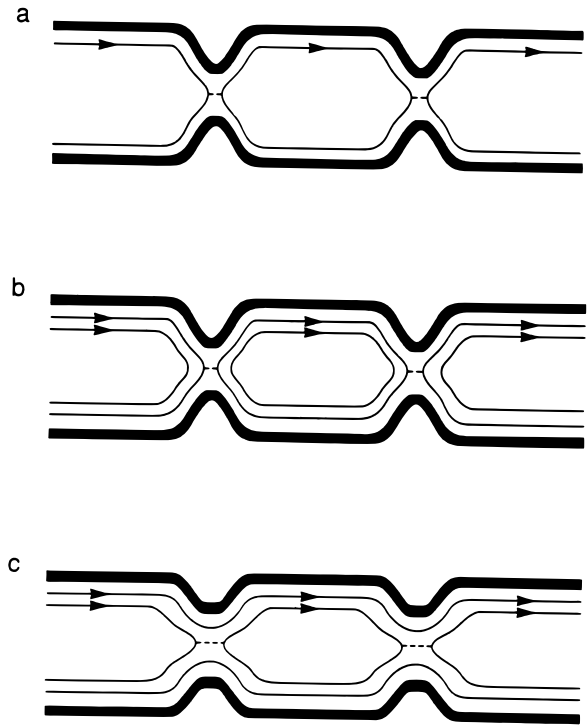}}
\caption{
Aharonov-Bohm magnetoconductance oscillations may occur due to resonant tunneling
through circulating edge states. Tunneling paths are indicated by dashed lines. (a) Only the first
Landau level is occupied. If the capacitance of the dot is sufficiently small, the Coulomb blockade
suppresses the Aharonov-Bohm oscillations. (b) Two Landau levels are occupied. Resonant tunneling
through the dot occurs predominantly through the first (outer) Landau level. The Aharonov-Bohm
effect is not suppressed by the charging energy. (c) Two Landau levels are occupied, one of which is
fully transmitted. Since the number of electrons in the dot is not discretized, no Coulomb blockade of
the Aharonov-Bohm effect is expected.
\label{fig17}
}
\end{figure}

Three cases of interest are illustrated in Fig.\ \ref{fig17}. In a strong magnetic field, only
edge states with $n=1$ corresponding to the first Landau level are occupied (Fig.\ \ref{fig17}a).
As the field is reduced, also the second Landau level, $n=2$, is occupied, as indicated
in Fig.\ \ref{fig17}b. Tunneling through the quantum dot still occurs predominantly through
the $n=1$ edge states, which have the largest tunnel probability through the barriers.
If the height $E_{\mathrm{B}}$ of the potential barriers is reduced, the $n=1$ edge states near the
Fermi level may have $E_{\mathrm{G}}>E_{\mathrm{B}}$, so that they form an extended edge channel. The edge
states with $n>1$ may still have $E_{\mathrm{G}}<E_{\mathrm{B}}$, and remain bound in the dot as before. As
illustrated in Fig.\ \ref{fig17}c, resonant tunneling now occurs predominantly through the edge
states belonging to the second Landau level.

\begin{figure}
\centerline{\includegraphics[width=8cm]{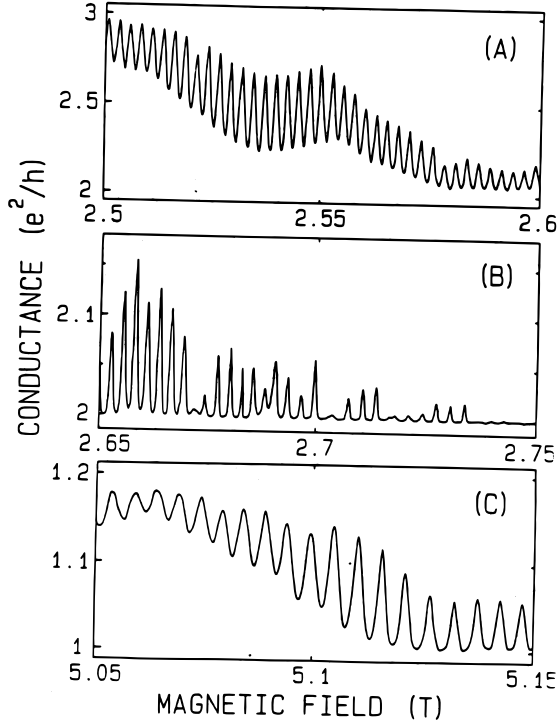}}
\caption{
Magnetoconductance of a quantum dot in the 2DEG of a GaAs-AlGaAs heterostructure of
1.5 $\mu \mathrm{m}$ diameter, with point contacts at entrance and exit serving as tunnel barriers. The temperature is
30 $\mathrm{m}\mathrm{K}$. (a) and (b) Aharonov-Bohm magnetoconductance oscillations due to resonant tunneling through
bound states belonging to the third (spin-split) edge channel. The first two (spin-split) Landau levels are
fully transmitted (cf.\ Fig.\ \ref{fig17}c). (c) Resonant tunneling through bound states belonging to the second
(spin-split) edge channel. The first (spin-split) edge channel is fully transmitted. (From Van Wees et
al.\cite{ref28})
\label{fig18}
}
\end{figure}

In the quantum Hall effect regime scattering between edge channels can be neglected
on length scales comparable to the diameter of the dot\cite{ref82} (this is known as adiabatic
transport\cite{ref8}). The edge channels may then be treated as independent parallel conduction
paths. The edge channels with $E_{\mathrm{G}}>E_{\mathrm{B}}$ contribute $e^{2}/h$ to the conductance. Resonant
tunneling through the edge states with $E_{\mathrm{G}}<E_{\mathrm{B}}$ gives an oscillating contribution to the
conductance of the quantum dot as a function of magnetic field. The periodicity of the
conductance oscillations can be deduced from the result (\ref{eq48}) for the edge state energy
spectrum. Resonant tunneling from the reservoir with Fermi energy $E_{\mathrm{F}}$ into an edge
state in the quantum dot is possible when $E_{\mathrm{F}}=E_{nm}$ for certain quantum numbers $n$
and $m$. For the edge states in the {\it n}-th Landau level the condition for resonant tunneling
is that the equipotential at the guiding center energy $E_{\mathrm{G}}\equiv E_{\mathrm{F}}-(n-\frac{1}{2})\hbar\omega_{\mathrm{c}}$ should
enclose $m+\gamma_{n}$ flux quanta, for some integer $m$. Let $A(B)$ denote the (magnetic field
dependent) area of the equipotential at energy $E_{\mathrm{G}}$. The {\it m}-th conductance peak occurs
at a magnetic field $B_{m}$ determined by $B_{m}A(B_{m})=(h/e)(m+\gamma_{n})$. The periodicity
$\Delta B\equiv B_{m+1}-B_{m}$ of the conductance oscillations from the {\it n}-th Landau level is obtained
by expanding $A(B)$ around $B_{m}$,
\be \Delta B=\frac{h}{e}[A(B_{m})+B_{m}A'(B_{m})]^{-1}
\equiv\frac{h}{e}\frac{1}{A_{\mathrm{eff}}(B_{m})}.   \label{eq49}
\ee
The effective area $A_{\mathrm{eff}}(B)$ can differ substantially from the geometrical area $A(B)$ in
the case of a smooth confining potential.\cite{ref28} The magnetoconductance oscillations are
approximately periodic in $B$ if the change in $A_{\mathrm{eff}}(B)$ in one period $\Delta B$ is much smaller
than the effective area itself. Since the change in $A_{\mathrm{eff}}$ is of order $h/eB$ per period,
while $A_{\mathrm{eff}}\sim mh/eB$, approximately periodic oscillations occur for $m\gg 1$. This is the
Aharonov-Bohm effect in the quantum Hall regime, first observed by Van Wees et al.\cite{ref28}
Their experimental results (reproduced in Fig.\ \ref{fig18}) correspond to the situation of
Fig.\ \ref{fig17}c with one (or more) fully transmitted edge channels.

\begin{figure}
\centerline{\includegraphics[width=8cm]{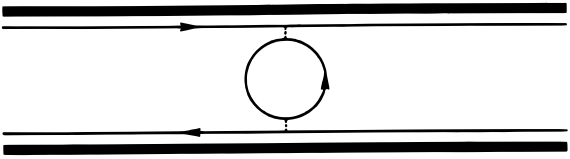}}
\caption{
A circulating edge state bound on a local potential maximum causes resonant backscattering,
thereby providing an alternative mechanism for Aharonov-Bohm magnetoconductance oscillations.
(From Beenakker and Van Houten.\cite{ref8})
\label{fig19}
}
\end{figure}

We close this subsection by mentioning that resonant backscattering (or resonant
reflection) can cause similar Aharonov-Bohm oscillations as those caused by resonant
transmission. Resonant backscattering may occur via a localized state bound on a potential maximum, created artificially (for example in a ring) or created by the presence
of disorder.\cite{ref83} The mechanism is illustrated in Fig.\ \ref{fig19}. Resonant backscattering leads
to a periodic suppression of the conductance, in contrast to the periodic enhancement
considered above.

\subsection{\label{sec4.2} Coulomb blockade of the Aharonov-Bohm effect}

Single-electron tunneling is governed by the transport of a single quantum of charge
$e$. The Aharonov-Bohm effect is governed by the flux quantum $h/e$. The present subsection addresses the interplay of these two quanta of nature in the integer quantum Hall
effect regime.

In the previous subsection we have discussed how resonant tunneling through circulating edge states can lead to magnetoconductance oscillations in a quantum dot with a
well-defined periodicity $\Delta B$, similar to the usual Aharonov-Bohm effect in a ring. There
is, however, an essential difference between the two geometries if only a single Landau
level is occupied.\cite{ref30} In each period $\Delta B$ the number of states below a given energy increases by one in a dot --- but stays constant in a ring. As a result, the Aharonov-Bohm
osclillations in the magnetoconductance of a quantum dot are accompanicd by an increase
of the charge of the dot by one elementary charge per period. That is of no consequence
if the Coulomb repulsion of the electrons can be neglected, but becomes important if
the dot has a small capacitance $C$ to the reservoirs, since then the electrostatic energy $e^{2}/C$ associated with the  incremental charging by single electrons has to be taken into
account.

Following Ref.\ \cite{ref30}, we analyze this problem by combining the results reviewed in
the previous sections. We apply Eq.\ (\ref{eq12}) to the energy spectrum shown in Fig.\ \ref{fig16}a. We consider here only the edge states from the lowest (spin-split) Landau level, so that the Aharonov-Bohm oscillations have a single periodicity. This corresponds to the strong-magnetic field limit. The magnetic field dependence of the edge states can be described approximately by a sequence of equidistant parallel lines,
\be
E_{p}={\rm constant}-\frac{\Delta E}{\Delta B}(B-p\Delta B),\label{eq50}
\ee
see Fig.\ \ref{fig16}a. For a circular quantum dot of radius $R$ with a hard-wall confining potential, one can estimate\cite{ref29} $\Delta B\sim h/eA$ and $\Delta E\sim\hbar\omega_{\rm c}l_{\rm m}/2R$. For a smooth confining potential $V(r)$ (with $l_{\rm m}V'\lesssim\hbar\omega_{\rm c}$) one has instead the estimates $\Delta B\sim(h/e)[A(B)+BA'(B)]^{-1}\sim(h/eA)[1-\hbar\omega_{\rm c}/RV'(R)]^{-1}$,\cite{ref28} and $\Delta E\sim h/\tau\sim l_{\rm m}^{2}V'(R)/R$, where $A(B)$ is the area enclosed by the equipotential of radius $R$ at the guiding center energy $V(R)=E-\frac{1}{2}\hbar\omega_{\rm c}$ (cf.\ Eq.\ (\ref{eq40}) for $n=1$). [The estimate for $\Delta E$ results from the correspondence between the level spacing and the period $\tau$ of the classical motion along the equipotential, with guiding-center-drift velocity $V'(R)/eB$.]

On substitution of Eq.\ (\ref{eq50}) into Eq.\ (\ref{eq12}), one finds the condition
\be
N\left(\Delta E+\frac{e^{2}}{C}\right)=\frac{\Delta E}{\Delta B}B_{N}+E_{\rm F}+{\rm constant}\label{eq51}
\ee
for the magnetic field value $B_{N}$ of the $N$-th conductance peak. The $B$-dependence of the reservoir Fermi energy can be neglected in Eq.\ (\ref{eq51}) in the case of a hard-wall confining potential (since ${\rm d}E_{\rm F}/{\rm d}B\approx\hbar\omega_{\rm c}/B\ll\Delta E/\Delta B$). The periodicity $B^{\ast}\equiv B_{N+1}-B_{N}$ of the Aharonov-Bohm oscillations is thus given by
\be
\Delta B^{\ast}=\Delta B\left(1+\frac{e^{2}}{C\Delta E}\right).\label{eq52}
\ee
[In the case of a smooth confining potential, the term $\Delta B$ in the enhancement factor of
Eq.\ (\ref{eq52}) should be replaced by the term $\Delta B[1+ (\Delta B/\Delta E)(\mathrm{d}E_{\mathrm{F}}/\mathrm{d}B)]^{-1}\sim h/eA$, under
the assumption that the Fermi energy in the reservoir is pinned to the lowest Landau
level, i.e.\ $E_{\mathrm{F}}=\frac{1}{2}\hbar\omega_{\mathrm{c}}$.] We conclude from Eq.\ (\ref{eq52}) that the charging energy enhances the
spacing of two subsequent peaks in $G$ versus $B$ by a factor $1+e^{2}/C\Delta E$. The periodicity
of the magnetoconductance oscillations is lost if $\Delta B^{*}$ becomes so large that the linear
approximation (\ref{eq50}) for $E_{p}(B)$ breaks down. Since Eq.\ (\ref{eq50}) holds at most over an energy
range of the Landau level separation $\hbar\omega_{\mathrm{c}}$, this suppression of the Aharonov-Bohm effect
occurs when $(\Delta E/\Delta B)\Delta B^{*}\gtrsim\hbar\omega_{\mathrm{c}}$, i.e.\ when $e^{2}/C\gtrsim\hbar\omega_{\mathrm{c}}$.

\begin{figure}
\centerline{\includegraphics[width=8cm]{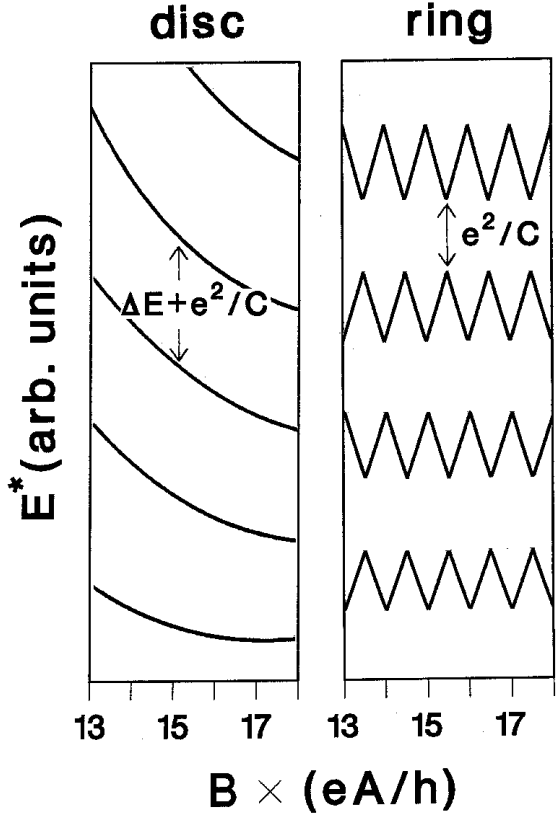}}
\caption{
Renormalized energy levels, defined by Eq.\ (\ref{eq12}), corresponding to the bare energy levels
shown in Fig.\ \ref{fig16}. (From Beenakker et al.\cite{ref30})
\label{fig20}
}
\end{figure}

\begin{figure}
\centerline{\includegraphics[width=8cm]{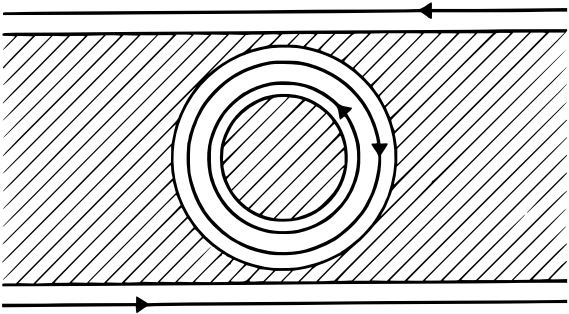}}
\caption{
Schematic layout of a semiconductor nanostructure proposed to demonstrate the Coulomb
blockade of the Aharonov-Bohm effect in a quantum dot, and its recovery upon transformation of the
device into a ring (by applying a negative voltage to the central gate). (From Beenakker et al.\cite{ref30})
\label{fig21}
}
\end{figure}

The Aharonov-Bohm oscillations with bare periodicity $\Delta B=h/eA$ are recovered if
one makes a hole in the disc, which is sufficiently large that the area $S$ of the conducting
region is much smaller than the enclosed area $A$. The inner perimeter of the resulting
ring supports a second set of edge states, which travel around the ring in opposite
direction as the first set of edge states at the outer perimeter. We compare in Fig.\ \ref{fig16} the
energy spectrum for a disc\cite{ref80} and a ring.\cite{ref81} The two sets of clockwise and counter-clockwise propagating edge states in a ring are distinguished by the opposite sign of
$\mathrm{d}E_{p}/\mathrm{d}B$, i.e.\ of the magnetic moment. Each set of edge states leads to oscillations in
the magnetoconductance of a ring with the same period $\Delta B$, but shifted in phase (and
in general with different amplitude, because the edge states at the inner perimeter have
a smaller tunneling probability to the reservoir than those at the outer perimeter). The
charging energy does not modify $\Delta B$ in a ring, because
\[
E_{p}(B)=E_{p}(B+\Delta B)\;\; ({\rm ring}).
\]
In a disc, in contrast, one has according to Eq.\ (\ref{eq50}),
\[
E_{p}(B)=E_{p+1}(B+\Delta B)\;\; ({\rm disc}).
\]
To illustrate the difference, we compare in Fig.\ \ref{fig20} for disc and ring the renormalized
energy levels $E_{p}^{*}$ [defined in Eq.\ (\ref{eq12})]. The effect of the charging energy in a ring is to
open an energy gap of magnitude $e^{2}/C$ in $E_{p}^{*}$. This gap will not affect the conductance
oscillations as a function of $B$ (at constant or slowly varying $E_{\mathrm{F}}$). A controlled experimental demonstration of the influence of Coulomb repulsion on the AB effect may be
obtained in a system which can be transformed from a disc into a ring. What we have
in mind is a geometry such as shown in Fig.\ \ref{fig21}, which has an additional gate within
the gates shaping the disc. By applying a negative voltage to this additional gate one
depletes the central region of the quantum dot, thereby transforming it into a ring. In
order to estimate the mutual capacitance $C$ between the undepleted quantum disc and
the adjacent 2DEG reservoirs, we note that only a circular strip of width $l_{\mathrm{m}}$ and radius
$R$ along the circumference of the disc contributes to $C$. The central region of the dot is
incompressible in the quantum Hall effect regime, and thus behaves as a dielectric as far
as the electrostatics is concerned. The capacitance $C$ contains contributions from the
self-capacitance of this strip as well as from its capacitance to the gate. (We assume that
the gate is electrically connected to the 2DEG reservoirs.) Both contributions are of order $\epsilon R$, with a numerical prefactor of order unity which depends only logarithmically on
the width of the strip and the separation to the gate\cite{ref57} ($\epsilon$ is the dielectric constant). A
dot radius of 1 $\mu \mathrm{m}$ yields a charging energy $e^{2}/C\simeq 1\,\mathrm{m}\mathrm{e}\mathrm{V}$ for $\epsilon\simeq 10\,\epsilon_{0}$. This exceeds the
level separation $\Delta E\simeq\hbar\omega_{\mathrm{c}}l_{\mathrm{m}}/2R\simeq 2\mathrm{x}10^{-5}\,\mathrm{e}\mathrm{V}(\mathrm{T}/B)$ at a field of a few T. A significant
increase of the frequency of the AB oscillations should thus be observable on depletion
of the central region of the dot, even for a relatively large radius of 1 $\mu \mathrm{m}$. To observe a
full suppression of the AB effect in a sub-micron disc with $e^{2}/C\gtrsim\hbar\omega_{\mathrm{c}}$, and its recovery
on transformation to a ring, would be an ultimate test of the theory\cite{ref30} reviewed here.

The difference between a ring and a disc disappears if more than a single Landau
level is occupied in the disc. This occurs in the upper-left-hand corner in Fig.\ \ref{fig16}a. The
energy spectrum in a disc now forms a mesh pattern which is essentially equivalent to
that in a ring (Fig.\ \ref{fig16}b). There is no Coulomb-blockade of the Aharonov-Bohm effect
in such a case,\cite{ref32} as discussed below.

\subsection{\label{sec4.3} Experiments on quantum dots}

\begin{figure}
\centerline{\includegraphics[width=8cm]{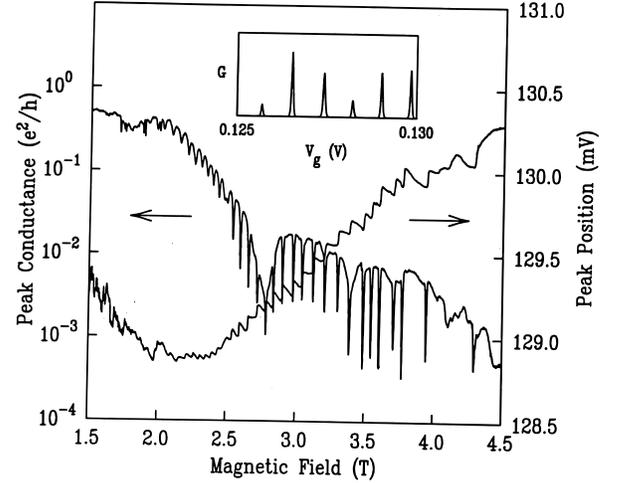}}
\caption{
Effect of a magnetic field on the height and position of a conductance peak in a quantum
dot in a GaAs-AlGaAs heterostructure, of the design shown in Fig.\ \ref{fig2}b. The temperature is 100 $\mathrm{m}\mathrm{K}$.
Inset: Coulomb-blockade oscillations as a function of gate voltage, for $B=3$ T. (From McEuen et al.\cite{ref32})
\label{fig22}
}
\end{figure}

We propose that the observation in a quantum dot of Aharonov-Bohm magnetoconductance oscillations by Van Wees et al.\cite{ref82} was made possible by the presence of one
or more extended edge channels, as in Fig.\ \ref{fig17}c (all of the succesful observations were,
to our knowledge, made for $G>e^{2}/h$). In the presence of extended states the charge on
the dot varies continuously, so that the Coulomb blockade of the Aharonov-Bohm effect
discussed above is not operative. A direct experimental test of this interpretation would
be desirable. This could be done by repeating the experiment in different magnetic field
regimes, both with and without the presence of an extended edge channel.

Even if the magnetoconductance oscillations are suppressed, it is still possible to
observe Coulomb-blockade oscillations in the conductance as a function of gate voltage
(at fixed magnetic field). Previous observations of conductance oscillations as a function
of gate voltage which were not observed as a function of $B$ have been attributed to the
Aharonov-Bohm effect,\cite{ref84,ref85} but might well have been Coulomb-blockade oscillations
instead.

An extended edge channel is one way to remove the Coulomb blockade of the
Aharonov-Bohm effect. A second circulating edge channel in the quantum dot is another way, exploited by McEuen et al.\cite{ref32} They observed conductance oscillations both
as a function of gate voltage and as a function of magnetic field in a quantum dot of
the design shown in Fig.\ \ref{fig2}b. Their main experimental results are reproduced in Fig.\ \ref{fig22}.
The trace of conductance versus gate voltage at $B=3\,\mathrm{T}$ (Fig.\ \ref{fig22}, inset) exhibits the
Coulomb-blockade oscillations, with an approximately constant periodicity. The main
curves in Fig.\ \ref{fig22} show that the height and position of a particular peak vary with $B$ in a
striking fashion. In the region between 2.5 and 3.5 $\mathrm{T}$ the peak height is periodically suppressed by as much as an order of magnitude, while the position of the peak oscillates
synchronously around a slowly varying background. In this field regime two Landau
levels are occupied in the dot, as in Fig.\ \ref{fig17}b, the lowest of which is spin-degenerate.

\begin{figure}
\centerline{\includegraphics[width=8cm]{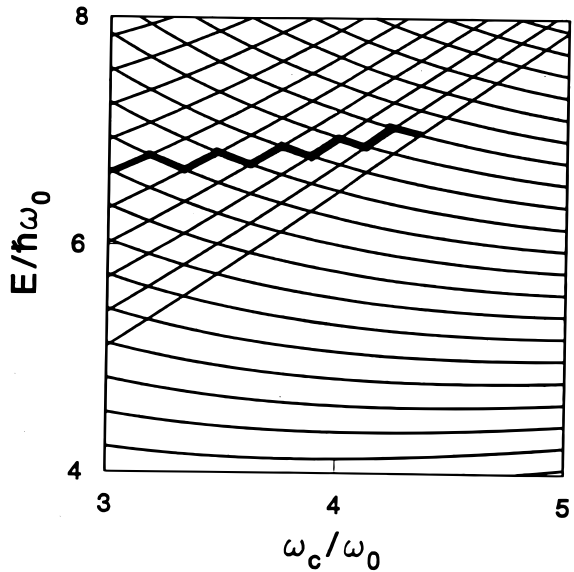}}
\caption{
Close-up of the energy spectrum of Fig.\ \ref{fig15} (after McEuen et al.\cite{ref32}). The heavy line indicates the energy of the highest occupied state for a fixed number (23) of electrons in the dot. In each period of the saw-tooth a single electron is transferred from the second Landau level (rising lines) to the first (falling lines).
\label{fig23}
}
\end{figure}

These observations have been explained by McEuen et al.\ in terms of the theory of Coulomb-blockade oscillations in the resonant tunneling regime. The one-electron energy spectrum in the range of two occupied Landau levels is shown in Fig.\ \ref{fig23} (for the case of a parabolic confining potential, cf.\ Fig.\ \ref{fig15}). The experiment is performed at 100 mK, which is presumably in the resonant tunneling regime $k_{\rm B}T<\Delta E$. Thus, only a single state participates in the conduction through the dot. As indicated in Fig.\ \ref{fig23} (heavy line), this state belongs alternatingly to the first and the second Landau level (corresponding, respectively, to the falling and rising line segments of the sawtooth in Fig.\ \ref{fig23}). Thus the tunnel rate into this state is alternatingly large and small. The periodic suppression of the peak height in Fig.\ \ref{fig22} reflects this difference in tunnel rates.

\begin{figure}
\centerline{\includegraphics[width=8cm]{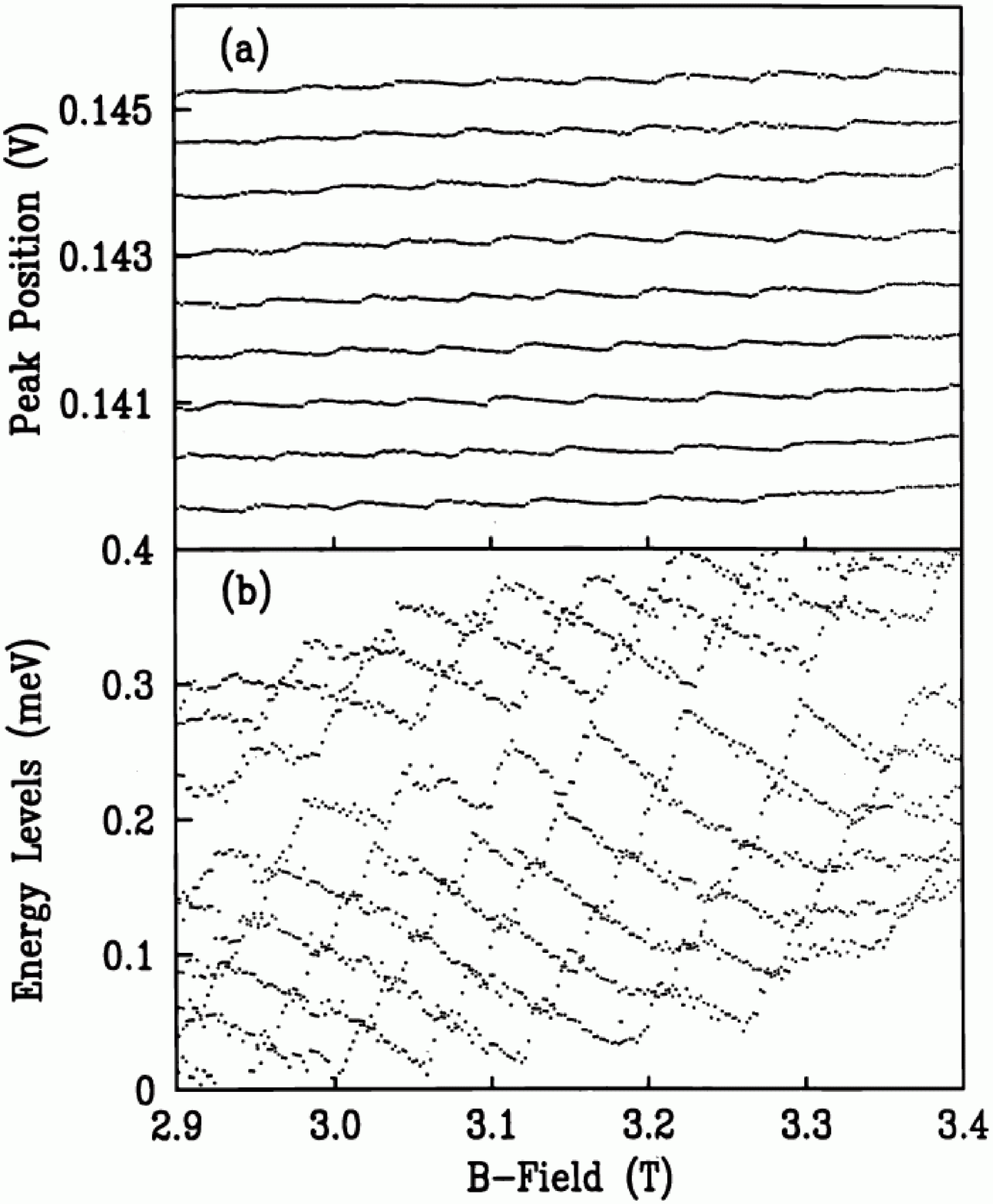}}
\caption{
(a) Peak position as a function of magnetic field for a series of consecutive Coulomb-blockade oscillations in a quantum dot with two occupied Landau levels. (b) Energy spectrum of the
dot obtained from the data in (a) after subtraction of the charging energy. (From McEuen et al.\cite{ref32})
\label{fig24}
}
\end{figure}

According to Eqs.\ (\ref{eq12}) and (\ref{eq14}), the gate voltage of the $N$-th peak shifts with $B$ according to
\be
\delta\phi_{\rm gate}=\frac{1}{\alpha e}\frac{\partial(E_{N}-E_{\rm F})}{\partial B}\delta B,\label{eq53}
\ee
with $\alpha$ defined in Eq.\ (\ref{eq17}). McEuen et al.\ determined $\alpha$ from the temperature dependence of the peak with, and neglected the change in $E_{\rm F}$ with $B$, as well as the difference between the electrostatic potential $\phi_{\rm gate}$ and the measured electrochemical potential $V_{\rm gate}$. The measured shift of the peak position with $B$ (see Fig.\ \ref{fig22}) then directly yields the shift in energy $E_{N}$. In this way they were able to map out the one-electron spectrum of the dot (Fig.\ \ref{fig24}). (To arrive at the bare energy spectrum a constant charging energy $e^{2}/C$ was subtracted for each consecutive level.) The similarity of Figs.\ \ref{fig23} and \ref{fig24}b is quite convincing. An unexplained effect is the gap in the spectrum around 0.2 meV. Also, the level spacing in the first Landau level (the vertical separation between the falling lines in Fig.\ \ref{fig24}b) appears to be two times smaller than that in the second Landau level (rising lines). Although this might be related to spin-splitting,\cite{ref32} we feel that it is more likely 
that the assumption of a magnetic-field independent $E_{\mathrm{F}}$ is not justified. If, as should
be expected, $E_{\mathrm{F}}$ is pinned to the second Landau level, then a proper correction for the
Fermi level shift with $B$ would lead to a clock-wise rotation of the entire level spectrum
in Fig.\ \ref{fig24}b around $(B,E)=(0,0)$. The agreement with the theoretical spectrum would
then improve.

\begin{figure}
\centerline{\includegraphics[width=8cm]{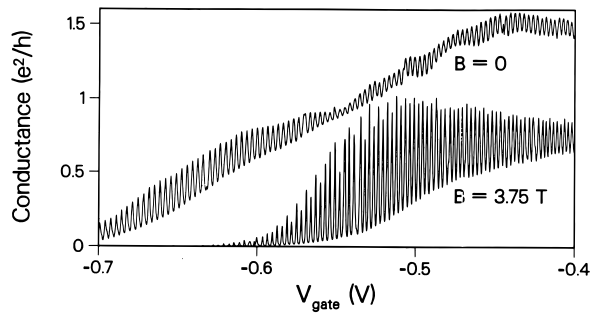}}
\caption{
Effect of a magnetic field on the conductance oscillations in a quantum dot in a GaAs-AlGaAs heterostructure, with a geometry as in Fig.\ \ref{fig2}c. The temperature is 50 $\mathrm{m}\mathrm{K}$. This is an effective
two-terminal conductance (obtained from a four-terminal conductance measurement, with the voltage
measured diagonally across the dot [Ref.\ \cite{ref8}, page 183].) (From Williamson et al.\cite{ref17})
\label{fig25}
}
\end{figure}

Coulomb-blockade oscillations as a function of gate voltage in the quantum Hall effect regime were studied by Williamson et al.\cite{ref17} in a quantum dot of the design shown
in Fig.\ \ref{fig2}c. They found that the amplitude of the oscillations was strongly enhanced compared to zero field, whereas the period was not much affected. (A similar enhancement of
the amplitude has been seen in disordered quantum wires, and possible explanations are
discussed below.) Representative traces of conductance versus gate voltage at zero field
and for $B=3.75\,\mathrm{T}$ are reproduced in Fig.\ \ref{fig25}. The oscillations in the presence of a field
are quite spectacular, of amplitude comparable to $e^{2}/h$. These experiments are in the
regime where the conductance of the individual barriers approaches $e^{2}/h$ as well, and
virtual tunneling processes may be important. Experimentally, the conductance minima are not exponentially suppressed (see Fig.\ \ref{fig25}), even though the temperature was
low $(100\,\mathrm{m}\mathrm{K})$. In addition, the conductance maxima in the zero-field trace exceed $e^{2}/h$.
These observations are also indicative of virtual tunneling processes.\cite{ref53,ref54} Finally, we
would like to draw attention to the slow beating seen in the amplitude of the oscillations
at zero field, which is suppressed at $B=3.75$ T. Instead, a weak doublet-like structure
becomes visible, reminiscent of that reported by Staring et al.\cite{ref12} for a disordered quantum wire in a strong magnetic field (see Fig.\ \ref{fig26}), discussed below. Further experimental
and theoretical work is needed to understand these intriguing effects of a magnetic field.

\subsection{\label{sec4.4} Experiments on disordered quantum wires}

\begin{figure}
\centerline{\includegraphics[width=8cm]{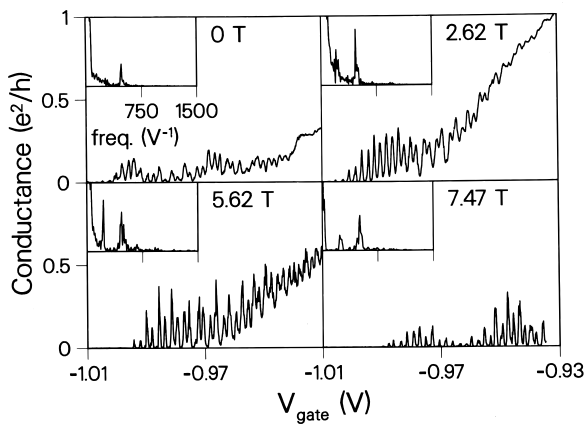}}
\caption{
Effect of a magnetic field on the Coulomb-blockade oscillations a disordered quantum wire
(as in Fig.\ \ref{fig11}), at 50 $\mathrm{m}\mathrm{K}$. Insets: Fourier transforms of the data, with the vertical axes of the curves at
0 $\mathrm{T}$ and 7.47 $\mathrm{T}$ magnified by a factor 2.5, relative to the curves at 2.62 $\mathrm{T}$ and 5.62 T. (From Staring et al.\cite{ref12})
\label{fig26}
}
\end{figure}

The effect of a parallel and perpendicular magnetic field on the conductance oscillations in a narrow channel in a Si inversion layer has been studied by Field et al.\cite{ref11}
Staring et al.\cite{ref12,ref13} investigated the effect of a perpendicular field on disordered
quantum wires in the 2DEG of a GaAs-AlGaAs heterostructure. Some of the data is
reproduced in Fig.\ \ref{fig26}. The Fourier transforms of the traces of conductance versus gate
voltage (insets) demonstrate a $B$-independent dominant frequency of 450 $\mathrm{V}^{-1}$. Curiously, as the magnetic field is increased a second peak in the Fourier transform emerges
at about half the dominant frequency. This second peak corresponds to an amplitude
modulation of the peaks, as is most clearly seen in the trace at 5.62 $\mathrm{T}$ where high and
low peaks alternate in a doublet-like structure. This feature is characteristic of this particular sample. Other channels showed different secondary effects, such as a much more
rapid oscillation superposed on the conductance trace for certain values of the magnetic
field.\cite{ref12} It is likely that the presence of additional segments in the wire plays a role.
The period $\delta V_{{\rm gate}}\sim 2.2\,\mathrm{m}\mathrm{V}$ of the dominant conductance oscillations is remarkably
insensitive to a strong magnetic field. Spin-splitting of the peaks was not observed, even
at the highest fields of 8 T. These qualitative observations agree with our interpretation
of the effect as Coulomb-blockade oscillations. In Sec.\ \ref{sec3.2} we have already had occasion to show that the temperature dependence of the lineshape of an isolated peak was
well accounted for by Eq.\ (\ref{eq24}), for a set of parameter values consistent with zero-field
experiments.

The height of the conductance peaks is enhanced by a field of intermediate strength
$(2\,\mathrm{T}<B<6\,\mathrm{T})$, followed by a decrease at stronger fields $(B\sim 7.5\,\mathrm{T})$. Also the width
of the peaks is reduced in a strong magnetic field. The largest isolated peaks (found
in a different sample\cite{ref13}) approach a height of $e^{2}/h$, measured two-terminally. A similar enhancement of the amplitude of the Coulomb-blockade oscillations by a magnetic
field was observed in a quantum dot\cite{ref17} (see Fig.\ \ref{fig25}). One explanation is that the
inelastic scattering rate is reduced by a magnetic field. In the low-temperature regime
$ k_{\mathrm{B}}T\lesssim h\Gamma$ this makes the peaks higher and narrower (cf.\ Sec.\ \ref{sec2.2}). In a disordered quantum wire the magnetic suppression of backscattering provides another mechanism for an
enhancement of the peak height because of the resulting reduction in series resistance.\cite{ref13}
Additionally, the modulation of the Fermi level in the quantum Hall effect regime
may lead to a non-monotonic variation with $B$ of the transmission probability $T(E_{\mathrm{F}})$,
and thus presumably of the tunnel rates $ h\Gamma$. The level degeneracy varies with $B$, becoming large when the Fermi energy coincides with a bulk Landau level in the dot. This
may also give rise to variations in the peak height.\cite{ref34} These are tentative explanations
of the surprising magnetic field dependence of the amplitude of the Coulomb-blockade
oscillations, which remains to be elucidated.

We close this subsection by noting that Staring et al.\cite{ref12} also measured magnetoconductance traces at fixed gate voltage. In contrast to the gate voltage scans, these
exhibited irregular structure only, with strong features corresponding to depopulation
of Landau levels. The absence of regular oscillations constitutes the first experimental
evidence for the predicted\cite{ref30} Coulomb blockade of the Aharonov-Bohm effect.

\acknowledgments

Valuable discussions with S. Colak, L. P. Kouwenhoven, N. C.
van der Vaart, J. G. Williamson, and the support of J. Wolter and M. F. H. Schuurmans
are gratefully acknowledged. Our experimental work has been made possible by C. T.
Foxon who has grown the necessary samples by molecular beam epitaxy, and by C. E.
Timmering who took care of the technology. We have benefitted from interactions with
the participants of the NATO ASI on Single Charge Tunneling. We thank our colleagues
at M.I.T., Delft, and Philips for their permission to reproduce some of their results. This
research was partly funded under the ESPRIT basic research action project 3133.

\appendix
\section{\label{appA} Conductance of a quantum dot coupled to two electron reservoirs}

\begin{figure}
\centerline{\includegraphics[width=8cm]{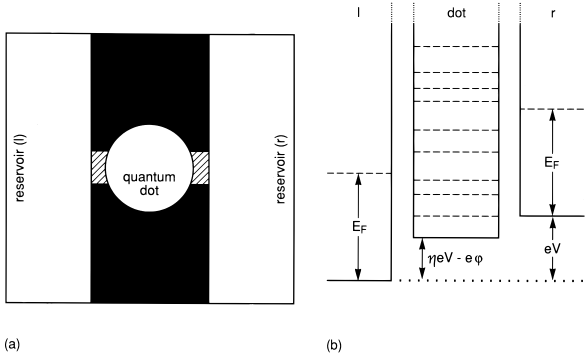}}
\caption{
(a) Schematic cross-section of the geometry studied in this appendix, consisting of a confined
region (``quantum dot'') weakly coupled to two electron reservoirs via tunnel barriers (hatched). (b)
Profile of the electrostatic potential energy (solid curve) along a line through the tunnel barriers. The
Fermi levels in the left and right reservoirs, and the discrete energy levels in the quantum dot are
indicated (dashed lines).
\label{fig27}
}
\end{figure}

Following the treatment by Beenakker,\cite{ref19} we derive in this appendix Eq.\ (\ref{eq24})
for the conductance of a confined region which is weakly coupled via tunnel barriers
to two electron reservoirs. The confined region, or ``quantum dot'', has single-electron
energy levels at $E_{p}$ $(p=1,2,$ $\ldots)$, labeled in ascending order and measured relative to
the bottom of the potential well. Each level contains either one or zero electrons. Spin
degeneracy can be included by counting each level twice, and other degeneracies can be
included similarly. Each reservoir is taken to be in thermal equilibrium at temperature $T$
and chemical potential $E_{\mathrm{F}}$. A continuum of states is assumed in the reservoirs, occupied
according to the Fermi-Dirac distribution
\be
f(E-E_{\mathrm{F}})=\left[1+\exp\left(\frac{E-E_{\mathrm{F}}}{kT}\right)\right]^{-1}. \label{eq54}
\ee
In Fig.\ \ref{fig27} we show schematically a cross-section of the geometry, and the profile of the
electrostatic potential energy along a line through the tunnel barriers.

A current $I$ can be passed through the dot by applying a potential difference $V$
between the two reservoirs. The tunnel rate from level $p$ to the left and right reservoirs in Fig.\ \ref{fig27} is denoted by $\Gamma_{p}^{\rm l}$ and $\Gamma_{p}^{\mathrm{r}}$, respectively. We assume that both $kT$ and
$\Delta E$ are $\gg h(\Gamma^{\rm l}+\Gamma^{\mathrm{r}})$ (for all levels participating in the conduction), so that the finite
width $h\Gamma=h(\Gamma^{\rm l}+\Gamma^{\mathrm{r}})$ of the transmission resonance through the quantum dot can be
disregarded. This assumption allows us to characterize the state of the quantum dot by
a set of occupation numbers, one for each energy level. (As discussed in Sec.\ \ref{sec2.2}, the
restriction $kT,$ $\Delta E\gg h\Gamma$ results in the conductance being much smaller than the quantum $e^{2}/h.$) We also assume conservation of energy in the tunnel process, thus neglecting
contributions of higher order in $\Gamma$ from tunneling via a {\it virtual\/} intermediate state in the quantum dot.\cite{ref53,ref53} We finally assume that inelastic scattering
takes place exclusively in the reservoirs --- not in the quantum dot. The effect of inelastic
scattering in the quantum dot is considered in Ref.\ \cite{ref19}.

Energy conservation upon tunneling from an initial state $p$ in the quantum dot
(containing $N$ electrons) to a final state in the left reservoir at energy $E^{\mathrm{f,l}}$ (in excess of
the local electrostatic potential energy), requires that
\be
E^{\mathrm{f,l}}(N)=E_{p}+U(N)-U(N-1)+\eta eV.   \label{eq55}
\ee
Here $\eta$ is the fraction of the applied voltage $V$ which drops over the left barrier. (As we
will see, this parameter $\eta$ drops out of the final expression for the conductance in linear
response.) The energy conservation condition for tunneling from an initial state $E^{\mathrm{i,l}}$ in
the left reservoir to a final state $p$ in the quantum dot is
\be
E^{\mathrm{i,l}}(N)=E_{p}+U(N+1)-U(N)+\eta eV,   \label{eq56}
\ee
where [as in Eq.\ (\ref{eq55})] $N$ is the number of electrons in the dot {\it before\/} the tunneling event.
Similarly, for tunneling between the quantum dot and the right reservoir one has the
conditions
\begin{eqnarray}
E^{\mathrm{f,r}}(N)&=&E_{p}+U(N)-U(N-1)-(1-\eta)eV,   \label{eq57}\\
E^{\mathrm{i,r}}(N)&=&E_{p}+U(N+1)-U(N)-(1-\eta)eV,   \label{eq58}
\end{eqnarray}
where $E^{\mathrm{i,r}}$ and $E^{\mathrm{f,r}}$ are the energies of the initial and final states in the right reservoir.

The stationary current through the left barrier equals that through the right barrier,
and is given by
\begin{eqnarray}
I&=&-e \sum_{p=1}^{\infty}\sum_{\{n_{i}\}}\Gamma_{p}^{\rm l}P(\{n_{i}\})\left(\delta_{n_{p},0}f(E^{\mathrm{i,l}}(N)-E_{\mathrm{F}})\right.\nonumber\\
&&\left.\mbox{}-\delta_{n_{p},1}[1-f(E^{\mathrm{f,l}}(N)-E_{\mathrm{F}})]\right).   \label{eq59}
\end{eqnarray}
The second summation is over all realizations of occupation numbers $\{n_{1},n_{2},\ldots\}\equiv
\{n_{i}\}$ of the energy levels in the quantum dot, each with stationary probability $P(\{n_{i}\})$.
(The numbers $n_{i}$ can take on only the values 0 and 1.) In equilibrium, this probability
distribution is the Gibbs distribution in the grand canonical ensemble:
\be
P_{\mathrm{e}\mathrm{q}}( \{n_{i}\})=\frac{1}{Z}\exp\left[-\frac{1}{kT}\left(\sum_{i=1}^{\infty}E_{i}n_{i}+U(N)-NE_{\mathrm{F}}\right)\right],   \label{eq60}
\ee
where $N\equiv\sum_{i}n_{i}$, and $Z$ is the partition function,
\be
Z=\sum_{\{n_{i}\}}\exp\left[-\frac{1}{kT}\left(\sum_{i=1}^{\infty}E_{i}n_{i}+U(N)-NE_{\mathrm{F}}\right)\right].   \label{eq61}
\ee

\begin{widetext}
The non-equilibrium probability distribution $P$ is a stationary solution of the kinetic
equation
\begin{eqnarray} 
\frac{\partial}{\partial t}P(\{n_{i}\})&=&0\nonumber\\
&=&-\sum_{p}P(\{n_{i}\})\delta_{n_{p},0}(\Gamma_{p}^{\rm l}f(E^{\mathrm{i,l}}(N)-E_{\mathrm{F}})+\Gamma_{p}^{\mathrm{r}}f(E^{\mathrm{i,r}}(N)-E_{\mathrm{F}}))\nonumber\\
&&\mbox{}- \sum_{p}P(\{n_{i}\})\delta_{n_{p},1}(\Gamma_{p}^{\rm l}[1-f(E^{\mathrm{f,l}}(N)-E_{\mathrm{F}})]+\Gamma_{p}^{\mathrm{r}}[1-f(E^{\mathrm{f,r}}(N)-E_{\mathrm{F}})])\nonumber\\
&&\mbox{}+ \sum_{p}P(n_{1},\ldots n_{p-1},1,n_{p+1},\ldots)\delta_{n_{p},0}(\Gamma_{p}^{\rm l}[1-f(E^{\mathrm{f,l}}(N+1)-E_{\mathrm{F}})]+\Gamma_{p}^{\mathrm{r}}[1-f(E^{\mathrm{f,r}})(N+1)-E_{\mathrm{F}})])\nonumber\\
&&\mbox{}+ \sum_{p}P(n_{1},\ldots n_{p-1},0,n_{p+1},\ldots)\delta_{n_{p},1}(\Gamma_{p}^{\rm l}f(E^{\mathrm{i,l}}(N-1)-E_{\mathrm{F}})+\Gamma_{p}^{\mathrm{r}}f(E^{\mathrm{i,r}}(N-1)-E_{\mathrm{F}})).   \label{eq62}
\end{eqnarray}
The kinetic equation (\ref{eq62}) for the stationary distribution function is equivalent to the set
of detailed balance equations (one for each $p=1,2,\ldots$)
\begin{eqnarray}
&&P(n_{1},\ldots n_{p-1},1,n_{p+1}, \ldots)(\Gamma_{p}^{\rm l}[1-f(E^{\mathrm{f,l}}(\tilde{N}+1)-E_{\mathrm{F}})]+\Gamma_{p}^{\mathrm{r}}[1-f(E^{\mathrm{f,r}}(\tilde{N}+1)-E_{\mathrm{F}})])\nonumber\\
&&=P(n_{1},\ldots n_{p-1},0,n_{p+1}, \ldots)(\Gamma_{p}^{\rm l}f(E^{\mathrm{i,l}}(\tilde{N})-E_{\mathrm{F}})+\Gamma_{p}^{\mathrm{r}}f(E^{\mathrm{i,r}}(\tilde{N})-E_{\mathrm{F}})),  \label{eq63}
\end{eqnarray}
with the notation $\tilde{N}\equiv\sum_{i\neq p}n_{i}$.

A similar set of equations formed the basis for the work of Averin, Korotkov, and
Likharev on the Coulomb staircase in the non-linear $I-V$ characteristic of a quantum dot.\cite{ref34}
To simplify the solution of the kinetic equation, they assumed that the charging
energy $e^{2}/C$ is much greater than the average level spacing $\Delta E$. In this chapter we
restrict ourselves to the regime of linear response, appropriate for the Coulomb-blockade
oscillations. Then the conductance can be calculated exactly and analytically.

The (two-terminal) linear response conductance $G$ of the quantum dot is defined as
$G=I/V$ in the limit $V\rightarrow 0$. To solve the linear response problem we substitute
\be
P(\{n_{i}\})\equiv P_{\mathrm{e}\mathrm{q}}(\{n_{i}\})\left(1+\frac{eV}{kT}\Psi(\{n_{i}\})\right)   \label{eq64}
\ee
into the detailed balance equation (\ref{eq63}), and linearize with respect to $V$. One finds
\begin{eqnarray}
&&P_{\mathrm{e}\mathrm{q}}(n_{1},\ldots n_{p-1},1,n_{p+1},\ldots)(\Psi(n_{1},\ldots n_{p-1},1,n_{p+1}, \ldots)(\Gamma_{p}^{\rm l}+\Gamma_{p}^{\mathrm{r}})[1-f(\epsilon)]-[\Gamma_{p}^{\rm l}\eta-\Gamma_{p}^{\mathrm{r}}(1-\eta)]kTf'(\epsilon))\nonumber\\
&&=P_{\mathrm{e}\mathrm{q}}(n_{1},\ldots n_{p-1},0,n_{p+1},\ldots)(\Psi(n_{1},\ldots n_{p-1},0,n_{p+1}, \ldots)(\Gamma_{p}^{\rm l}+\Gamma_{p}^{\mathrm{r}})f(\epsilon)+[\Gamma_{p}^{\rm l}\eta-\Gamma_{p}^{\mathrm{r}}(1-\eta)]kTf'(\epsilon)),   \label{eq65}
\end{eqnarray}
where $ f'(\epsilon)\equiv df(\epsilon)/d\epsilon$, and we have abbreviated $\epsilon\equiv E_{p}+U(\tilde{N}+1)-U(\tilde{N})-E_{\mathrm{F}}$.

Equation (\ref{eq65}) can be simplified by making subsequently the substitutions
\begin{eqnarray}
&&1- f(\epsilon)=f(\epsilon)\mathrm{e}^{\epsilon/kT}, \label{eq66}\\
&&P_{\mathrm{e}\mathrm{q}}(n_{1},\ldots n_{p-1},1,n_{p+1},\ldots)=P_{\mathrm{e}\mathrm{q}}(n_{1}, \ldots n_{p-1},0,n_{p+1},\ldots)\mathrm{e}^{-\epsilon/kT},   \label{eq67}\\
&&kTf'(\epsilon)(1+\mathrm{e}^{-\epsilon/kT})=-f(\epsilon).   \label{eq68}
\end{eqnarray}
The factors $P_{\mathrm{e}\mathrm{q}}$ and $f$ cancel, and one is left with the simple equation
\be
\Psi(n_{1},\ldots n_{p-1},1,n_{p+1},\ldots)=\Psi(n_{1},\ldots n_{p-1},0,n_{p+1}, \ldots)+\frac{\Gamma_{p}^{\mathrm{r}}}{\Gamma_{p}^{\rm l}+\Gamma_{p}^{\mathrm{r}}}-\eta.   \label{eq69}
\ee
The solution is
\be
\Psi(\{n_{i}\})={\rm constant}+ \sum_{i=1}^{\infty}n_{i}\left(\frac{\Gamma_{i}^{\mathrm{r}}}{\Gamma_{i}^{\rm l}+\Gamma_{i}^{\mathrm{r}}}-\eta\right). \label{eq70}
\ee
The constant first term in Eq.\ (\ref{eq70}) takes care of the normalization of $P$ to first order
in $V$, and need not be determined explicitly. Notice that the first order non-equilibrium
correction $\Psi$ to $P_{\mathrm{e}\mathrm{q}}$ is {\it zero\/} if $\eta=\Gamma_{i}^{\rm r}/(\Gamma_{i}^{\rm l}+\Gamma_{i}^{\mathrm{r}})$ for all $i$. This will happen in particular
for two identical tunnel barriers (when $\eta=\frac{1}{2}$, $\Gamma_{i}^{\rm l}=\Gamma_{i}^{\mathrm{r}}$). Because of the symmetry of the
system, the distribution function then contains only terms of {\it even\/} order in $V$.

Now we are ready to calculate the current $I$ through the quantum dot to first order
in $V$. Linearization of Eq.\ (\ref{eq59}), after substitution of Eq.\ (\ref{eq64}) for $P$, gives
\begin{eqnarray}
I&=&-e \frac{eV}{kT}\sum_{p}\sum_{\{n_{i}\}}\Gamma_{p}^{\rm l}P_{\mathrm{e}\mathrm{q}}(\{n_{i}\})(\delta_{n_{p},0}\eta kTf'(\epsilon)+\delta_{n_{p},1}\eta kTf'(\epsilon)\nonumber\\
&&\mbox{}+\Psi(\{n_{i}\})\delta_{n_{p},0}f(\epsilon)-\Psi(\{n_{i}\})\delta_{n_{p},1}[1-f(\epsilon)])\nonumber\\
&=& \frac{e^{2}V}{kT}\sum_{p}\sum_{\{n_{i}\}}\Gamma_{p}^{\rm l}P_{\mathrm{e}\mathrm{q}}(\{n_{i}\})\delta_{n_{p},0}f(E_{p}+U(N+1)-U(N)-E_{\mathrm{F}})\nonumber\\
&&\mbox{}\times[\eta+\Psi(n_{1},\ldots n_{p-1},1,n_{p+1},\ldots)-\Psi(n_{1},\ldots n_{p-1},0, n_{p+1},\ldots)]\nonumber\\
&=& \frac{e^{2}V}{kT}\sum_{p}\sum_{\{n_{i}\}}\frac{\Gamma_{p}^{\rm l}\Gamma_{p}^{\mathrm{r}}}{\Gamma_{p}^{\rm l}+\Gamma_{p}^{\mathrm{r}}}P_{\mathrm{e}\mathrm{q}}(\{n_{i}\})\delta_{n_{p},0}f(E_{p}+U(N+1)-U(N)-E_{\mathrm{F}}).   \label{eq71}
\end{eqnarray}
In the second equality we have again made use of the identities (\ref{eq66})--(\ref{eq68}), and in the
third equality we have substituted Eq.\ (\ref{eq69}). Notice that the parameter $\eta$ has dropped
out of the final expression for $I$.

We define the equilibrium probability distributions
\begin{eqnarray}
P_{\mathrm{e}\mathrm{q}}(N)&=& \sum_{\{n_{i}\}}P_{\mathrm{e}\mathrm{q}}(\{n_{i}\})\delta_{N,\sum_{i}n_{i}}=\frac{\exp(-\Omega(N)/kT)}{\sum_{N}\exp(-\Omega(N)/kT)},   \label{eq72}\\
F_{\mathrm{e}\mathrm{q}}(E_{p}|N)&=& \frac{1}{P_{\mathrm{e}\mathrm{q}}(N)}\sum_{\{n_{i}\}}P_{\mathrm{e}\mathrm{q}}(\{n_{i}\})\delta_{n_{p},1}\delta_{N,\sum_{i}n_{i}}\nonumber\\
&=& \exp({\cal F}(N)/kT)\sum_{\{n_{i}\}}\exp\left(-\frac{1}{kT}\sum_{i=1}^{\infty}E_{i}n_{i}\right)\delta_{n_{p},1}\delta_{N,\sum_{i}n_{i}}. \label{eq73}
\end{eqnarray}
\end{widetext}
Here $\Omega(N)$ is the thermodynamic potential of the quantum dot, and ${\cal F}(N)$ is the free
energy of the internal degrees of freedom:
\begin{eqnarray}
&&\Omega(N)={\cal F}(N)+U(N)-NE_{\mathrm{F}}, \label{eq74}\\
&&{\cal F}(N)=-kT\ln\left[\sum_{\{n_{i}\}}\exp\left(-\frac{1}{kT}\sum_{i=1}^{\infty}E_{i}n_{i}\right)\delta_{N,\sum_{i}n_{i}}\right].\nonumber\\
&&\label{eq75}
\end{eqnarray}
The function $P_{\mathrm{e}\mathrm{q}}(N)$ is the probability that the quantum dot contains $N$ electrons in
equilibrium; The function $F_{\mathrm{e}\mathrm{q}}(E_{p}|N)$ is the conditional probability in equilibrium that
level $p$ is occupied given that the quantum dot contains $N$ electrons. In terms of these
distribution functions, the conductance $G=I/V$ resulting from Eq.\ (\ref{eq71}) equals
\begin{eqnarray}
G&=& \frac{e^{2}}{kT}\sum_{p=1}^{\infty}\sum_{N=0}^{\infty}\frac{\Gamma_{p}^{\rm l}\Gamma_{p}^{\mathrm{r}}}{\Gamma_{p}^{\rm l}+\Gamma_{p}^{\mathrm{r}}}P_{\mathrm{e}\mathrm{q}}(N)[1-F_{\mathrm{e}\mathrm{q}}(E_{p}|N)]\nonumber\\
&&\mbox{}\times f(E_{p}+U(N+1)-U(N)-E_{\mathrm{F}}). \label{eq76}
\end{eqnarray}
In view of Eqs.\ (\ref{eq66}) and (\ref{eq67}), Eq.\ (\ref{eq76}) can equivalently be written in the form
\begin{eqnarray}
G&=& \frac{e^{2}}{kT}\sum_{p=1}^{\infty}\sum_{N=1}^{\infty}\frac{\Gamma_{p}^{\rm l}\Gamma_{p}^{\mathrm{r}}}{\Gamma_{p}^{\rm l}+\Gamma_{p}^{\mathrm{r}}}P_{\mathrm{e}\mathrm{q}}(N)F_{\mathrm{e}\mathrm{q}}(E_{p}|N)\nonumber\\
&&\mbox{}\times[1-f(E_{p}+U(N)-U(N-1)-E_{\mathrm{F}})].\nonumber\\
&&\label{eq77}
\end{eqnarray}
Redefining $P_{\mathrm{e}\mathrm{q}}(N)F_{\mathrm{e}\mathrm{q}}(E_{p}|N)=P_{\mathrm{e}\mathrm{q}}(N,n_{p}=1)$ we find Eq.\ (\ref{eq24}) as it appears in Sec.\ \ref{sec2.2}.

\end{document}